\documentclass[onecolumn,10pt,a4paper,sort&compress,aps,pra,showpacs,superscriptaddress]{revtex4-2}

\usepackage[english]{babel}

\usepackage[caption=false]{subfig}

\addto\captionsenglish{\renewcommand{\figurename}{FIG.}}

\usepackage{overpic}
\usepackage{amsmath,amssymb,amsthm,mathtools}
\usepackage{commath}
\usepackage{graphicx}
\usepackage{bm}
\usepackage{hyperref}
\usepackage[capitalise,nameinlink]{cleveref}
\newcommand{\aref}[1]{\hyperref[#1]{Appendix}}
\usepackage{etoolbox} 
\usepackage{siunitx}

\usepackage[normalem]{ulem}
\usepackage{comment}
\usepackage[usenames,dvipsnames]{color}

\renewcommand{\vec}[1]{\bm{#1}}
\newcommand{\intd}{\mathop{}\!\mathrm{d}}

\newcommand{\diff}[2]{\frac{\mathrm{d}{#1}}{\mathrm{d}{#2}}}
\newcommand{\pdiff}[2]{\frac{\partial{#1}}{\partial{#2}}}

\newcommand{\x}{\vec{x}}
\newcommand{\y}{\vec{y}}

\newcommand{\bigO}[1]{\mathit{O}(#1)}
\newcommand{\smallO}[1]{\mathit{o}(#1)}
\newcommand{\eff}[1]{#1_e}
\newcommand{\avg}[1]{\left\langle #1 \right\rangle}

\newcommand{\Ic}{I_c}
\newcommand{\Is}{I_s}
\newcommand{\Ix}{I_x}
\newcommand{\Iy}{I_y}
\newcommand{\Iz}{I_z}
\newcommand{\Iphi}{I_{\phi}}
\newcommand{\Gzeta}{G_{\zeta}}
\newcommand{\Gphi}{G_{\phi}}

\newcommand{\trans}[1]{#1_{P}}
\newcommand{\rot}[1]{#1_{A}}

\newcommand{\znew}{\tilde{z}}

\newcommand{\leishmex}{\textit{Leishmania mexicana}}

\begin{document}


\title{The effects of rapid yawing on simple swimmer models and planar Jeffery's orbits}

\author{Benjamin J. Walker}
\email{Corresponding author: benjamin.walker@maths.ox.ac.uk}
\affiliation{Mathematical Institute, University of Oxford, Oxford, OX2 6GG, UK}

\author{Kenta Ishimoto}
\email{ishimoto@kurims.kyoto-u.ac.jp}
\affiliation{Research Institute for Mathematical Sciences, Kyoto University, Kyoto, 606-8502, Japan}

\author{Eamonn A. Gaffney}
\email{gaffney@maths.ox.ac.uk}
\affiliation{Mathematical Institute, University of Oxford, Oxford, OX2 6GG, UK}

\author{Cl\'{e}ment Moreau}
\email{cmoreau@kurims.kyoto-u.ac.jp}
\affiliation{Research Institute for Mathematical Sciences, Kyoto University, Kyoto, 606-8502, Japan}

\author{Mohit P. Dalwadi}
\email{dalwadi@maths.ox.ac.uk}
\affiliation{Mathematical Institute, University of Oxford, Oxford, OX2 6GG, UK}


\date{\today}

\begin{abstract}
Over a sufficiently long period of time, or from an appropriate distance, the motion of many swimmers can appear smooth, with their trajectories appearing almost ballistic in nature and slowly varying in character. These long-time behaviours, however, often mask more complex dynamics, such as the side-to-side snakelike motion exhibited by spermatozoa as they swim, propelled by the frequent and periodic beating of their flagellum. Many models of motion neglect these effects in favour of smoother long-term behaviours, which are often of greater practical interest than the small-scale oscillatory motion. Whilst it may be tempting to ignore any yawing motion, simply assuming that any effects of rapid oscillations cancel out over a period, a precise quantification of the impacts of high-frequency yawing is lacking. In this study, we systematically evaluate the long-term effects of general high-frequency oscillations on translational and angular motion, cast in the context of microswimmers but applicable more generally. Via a multiple-scales asymptotic analysis, we show that rapid oscillations can cause a long-term bias in the average direction of progression. We identify sufficient conditions for an unbiased long-term effect of yawing, and we quantify how yawing modifies the speed of propulsion and the effective hydrodynamic shape when in shear flow. Furthermore, we investigate and justify the long-time validity of the derived leading-order solutions and, by direct computational simulation, we evidence the relevance of the presented results to a canonical microswimmer.
\end{abstract}

\maketitle

\section{Introduction}\label{sec:intro}
Swimming in nature is often driven by undulatory deformations, from the snake-like motion of spermatozoa at low Reynolds number to the inertia-dominated movement of fish and aquatic mammals \cite{Gazzola2014}. When viewed over a sufficiently long timeframe, or from an adequate distance, the motion of many of these self-propelled objects can appear smooth, with their trajectories appearing almost ballistic in nature and slowly varying in character. These long-term behaviours, however, often mask the details of complex short-term dynamics that arise from their relatively high-frequency undulation. Though it is tempting to neglect the effects of yawing over long-enough timescales, rationalising this by assuming that the effects of rapid oscillations cancel over a period, a precise investigation of the impacts of high-frequency yawing on the long-term behaviour of a swimmer or particle is lacking, even for the simplest models. Indeed, the nonlinear angular dependence in some simple models of motion, such as that of an active particle with constant speed and prescribed orientation, suggests that the effects of rapid yawing may not simply average out to zero, as might be expected in linear systems. Hence, in this study, we seek to systematically evaluate the long-term effects of simple yet general oscillations on translational and angular motion. Whilst we will frame much of this exploration in the context of both self-propelled and passive bodies in a Stokesian fluid, much of the analysis, particularly that of \cref{sec:trans}, applies more generally to active particle models.

\begin{figure}
    \centering
    \raisebox{-0.5\height}{\begin{overpic}[width=0.3\textwidth]{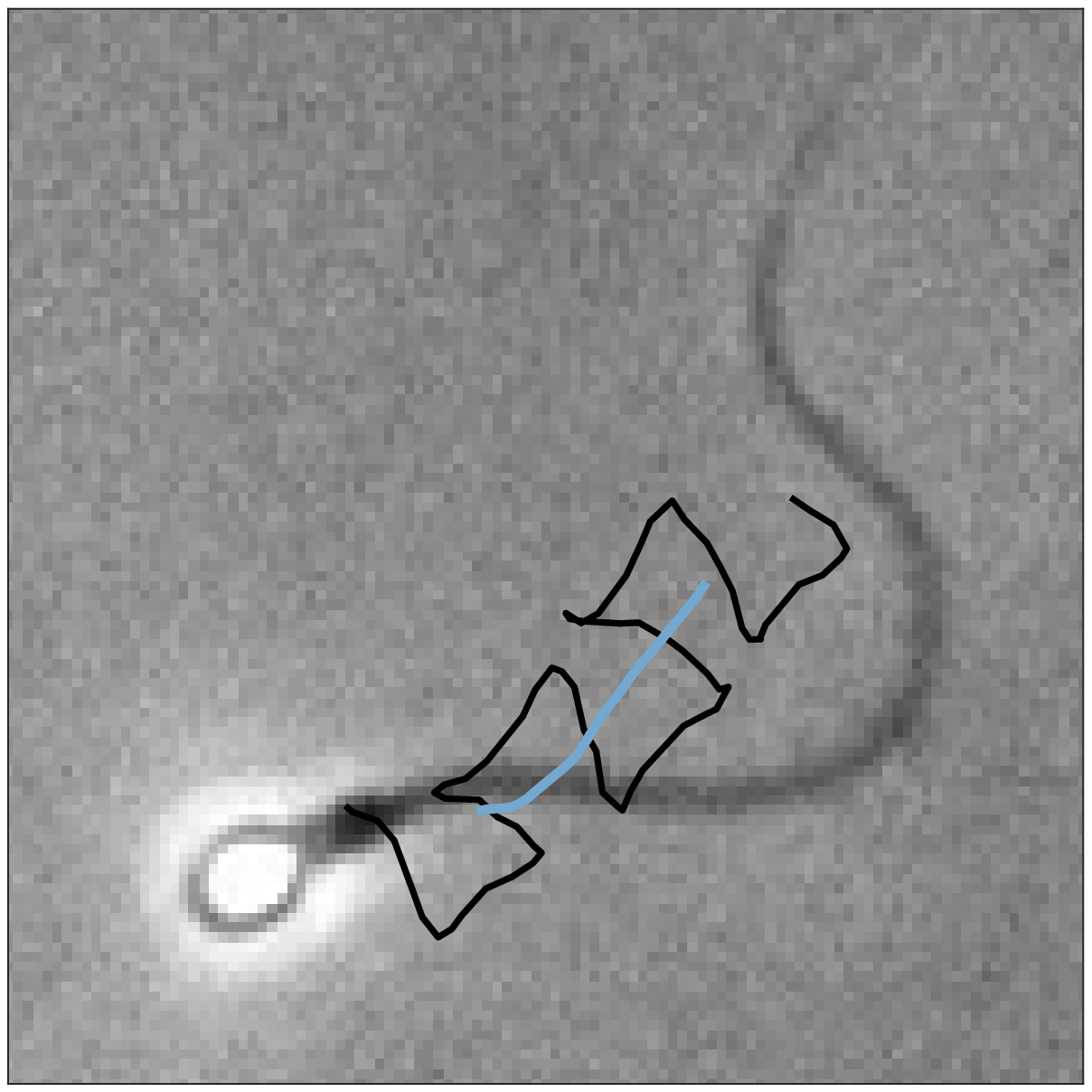}
    \put(-5,105){(a)}
    \end{overpic}}
    \hspace*{6em}
    \raisebox{-0.5\height}{\begin{overpic}[width=0.4\textwidth]{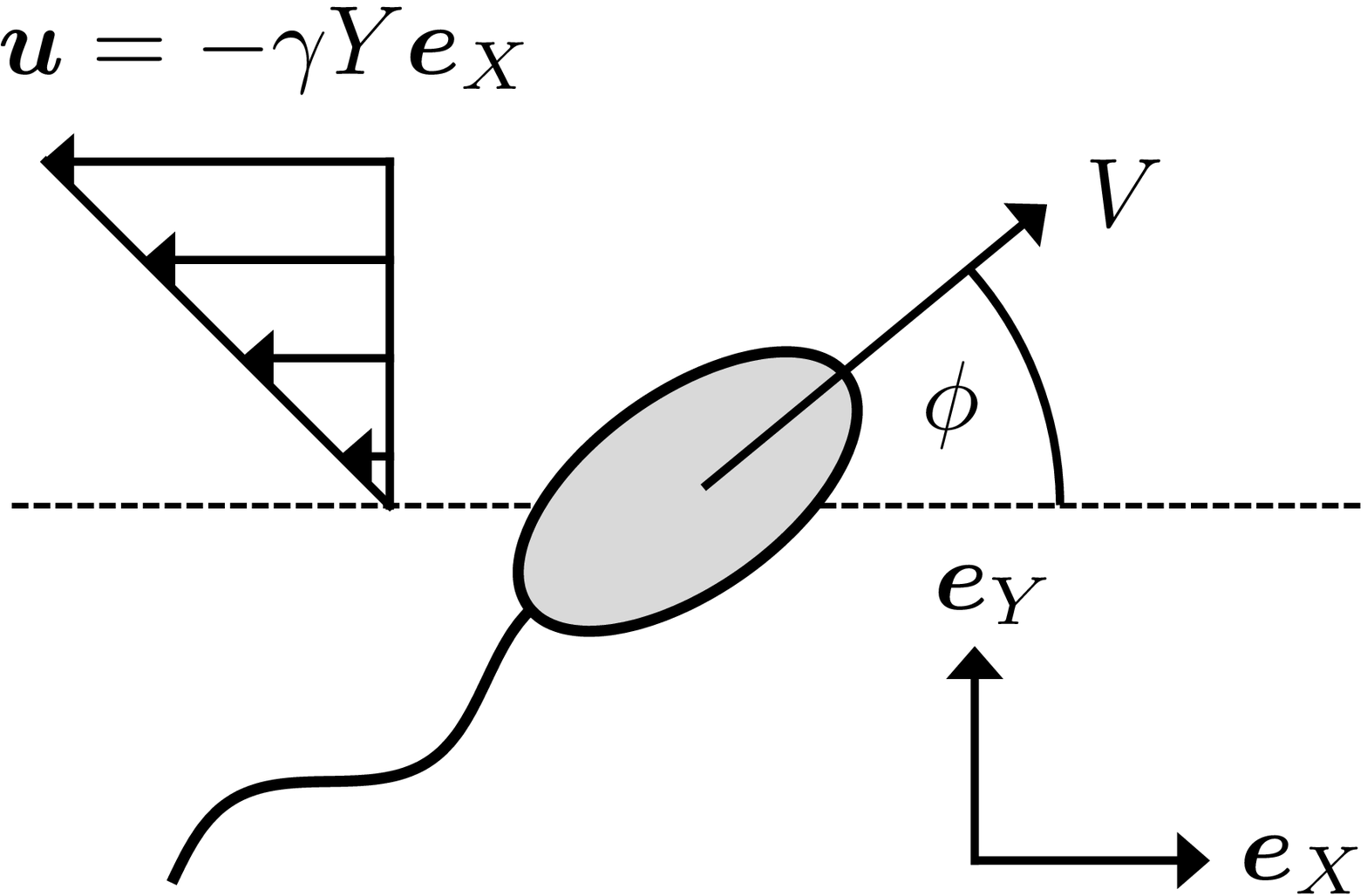}
    \put(-10,72){(b)}
    \end{overpic}}
    \caption{The path of a spermatozoon and a model self-propelled swimmer in flow. (a) The digitally tracked path of a human spermatozoon, shown as a black curve, with the smoothed average path shown in blue. (b) We illustrate an idealised swimmer, moving with speed $V$ at an orientation $\phi$ from the laboratory $\vec{e}_X$ axis. Though we show a model flagellated swimmer in shear flow, with $\vec{u}=-\gamma Y\vec{e}_X$, this setup is also applicable in the absence of background fluid flow, equivalent to taking $\gamma=0$, and to general self-propelled particles. Videomicroscopy from Movie 1 of the Supplemental Material of \citet{Ishimoto2017b}.}
    \label{fig:sperm_and_setup}
\end{figure}

In the well-explored context of microswimming, there are many examples of yawing bodies, with a multitude of swimmers being propelled by rapid oscillatory drivers. For instance, canonical amongst microswimmers is the aforementioned spermatozoon, with a slender flagellum providing the propulsive force that gives rise to its net head-first motion. Due to the progression of a travelling wave of displacement along their flagellum, as is typical (though not ubiquitous) for spermatozoa in common media \cite{Dresdner1981,Smith2009a,Hyakutake2019}, these cells undergo large side-to-side yawing, giving rise to complex oscillatory trajectories on a fast timescale (see \cref{fig:sperm_and_setup}a), though their long-term behaviours nevertheless appear to be smooth. In the study of this swimmer, along with that of many others, a broad variety of methods have been employed of varying degrees of complexity, ranging from the simplest phenomenological models, as we will consider in this work and are particularly common in the large-scale study of active matter \cite{Shaebani2020,Chepizhko2020, Bar2020}, to the most computationally and geometrically intricate \cite{pozrikidis2002,Smith2009,Walker2019,Shum2010,Rostami2019,Taketoshi2020}. Whilst the latter class of methods inherently includes the effects of rapid yawing in their formulation, it being an emergent property of geometrically faithful study, many simpler models neglect these effects in favour of capturing the smoother, long-term behaviours, which are often of greater interest than the rapid transient motion in applications \cite{Lauga2009,Ishimoto2014, Walker2019e,Shaebani2020,Ishimoto2018,Creppy2015,Dunkel2013,Tung2017,Woolley2003,Woolley2009,Shum2015}. However, despite the success of simple long-term or averaged models, it is not clear that the oft-seen neglect of the yawing motion exhibited by these swimmers is justified. Hence, the primary aim of this study will be to conduct a detailed investigation into the effects of rapid yawing on simple swimmer models, motivated by the valuable utility and persisting prevalence of such minimal representations.

A common simplifying feature of these models is to assume symmetric swimmer geometry. 
These geometrically coarse models have been widely employed in many contexts, from large-scale investigations of active matter to single-swimmer studies of guidance and control \cite{Rosser2014,Richards2016,Moreau2021,Moreau2021b,Somka2020,Saintillan2008,Kim2010}. Often, the reduction in geometrical complexity affords many advantages, such as numerical efficiency and analytical tractability. For example, a recent in silico study of \leishmex{} evidenced that an ellipsoid can be an appropriate model for this flagellated swimmer in the presence of a background shear flow \cite{Walker2018}, enabling rapid simulation in comparison to the computationally expensive boundary element method from which it was derived. Ellipsoid models in particular afford the significant advantage of having been the subject of much classical and recent study, with their behaviour in shear flow being the topic of \citeauthor{Jeffery1922}'s seminal work of \citeyear{Jeffery1922} \cite{Jeffery1922}. Since then, \citeauthor{Jeffery1922}'s analytical results, which we partially recount in \cref{sec:rot}, have been extended and generalised \cite{Bretherton1962,Brenner1964,Ishimoto2020b,Ishimoto2020a}. In particular, the work of \citet{Bretherton1962} and the recent contribution of \citet{Ishimoto2020a} considered broader classes of rigid bodies, including surfaces of revolution and those with particular symmetries. Though our understanding of the behaviours of such geometries in flow benefits significantly from this mature body of study, the impacts of rapid yawing on even the classical results of Jeffery's formulation remain unknown, despite the noted prevalence of rapid oscillations in microswimmers and the broad utilisation of simple models to study them. Thus, as a further aim of this work, we will investigate how, if at all, high-frequency yawing modifies the motion of particles admitted by Jeffery's theory and its subsequent generalisations for symmetric bodies.

Hence, we proceed by first considering a paradigm model of motion for a self-propelled particle in \cref{sec:trans}, rendered non-trivial by a prescribed oscillatory orientation that phenomenologically captures the features of rapid yawing. We will derive asymptotic solutions by exploiting the separated timescales of the problem, which relate to translation and rotation. This will allow us to understand the structure of the multiple-scales analysis we use throughout this paper, and to understand several basic implications of rapid yawing. Next, focusing on the case of planar rotations in a background shear flow, we will extend our analysis by introducing an appropriately coupled equation of motion for the rotational dynamics, seeking leading-order solutions in \cref{sec:rot}. From numerical validation of these leading-order solutions for both translational and angular motion, we will note that the asymptotic solution is surprisingly accurate, even on very long timescales. Motivated by this, we will further seek to establish sufficient conditions for such persistent validity by conducting a higher-order analysis in \cref{sec:drift}.

\section{A self-propelled, yawing swimmer}\label{sec:trans}
We begin by studying a minimal model of a self-propelled yawing swimmer in the plane, in the absence of any background flow. In particular, we will consider a body moving at a constant speed $V$ in an unbounded domain at a time-dependent angle $\phi(t)$ from a fixed laboratory axis (see \cref{fig:sperm_and_setup}b), denoting time by $t$. Although we will later generalise several of these simplifications, including an extension to three-dimensional dynamics in \cref{app:3D_trans}, and the effect of a background shear flow coupled to  angular motion in \cref{sec:rot}, we first present the simple two-dimensional problem in the absence of flow. This will allow us to understand the essential structure of the problem and the approach, including several implications of rapid yawing. Here and hereafter, we consider all quantities to be dimensionless. Whilst this entails that we may take $V=1$, without loss of generality, we will retain the symbol $V$ so as to clearly identify the role of the swimming speed in the analysis that follows.

With the coordinates of the swimmer position written as $(x(t),y(t))$ relative to laboratory unit vectors $\vec{e}_X$ and $\vec{e}_Y$, the swimmer motion is governed by the following ordinary differential equations (ODEs):
\begin{subequations}
\label{eq: simplest ODEs}
\begin{align}
    \diff{x}{t} &= V\cos{\phi}\,, \quad x(0) = 0,\\
    \diff{y}{t} &= V\sin{\phi}\,, \quad y(0) = 0,
\end{align}
\end{subequations}
measuring the orientation of $\phi(t)$ anticlockwise from the $X$-axis. We note that the initial conditions are chosen for convenience, and may be readily substituted. Here, we will seek to ascertain the effects of a rapidly oscillating $\phi$, modelling the yawing dynamics exhibited by many microswimmers, such as the flagellated spermatozoon or \leishmex{}.

\subsection{Sinusoidal yawing}\label{sec:trans:sine}
Though we later consider general yawing motions, it is instructive to first prescribe simple, sinusoidal oscillations in the orientation $\phi$, seeking to develop our intuition for the contributions of rapid angular variation. We denote the angular frequency of the oscillations by $\omega \gg 1$, which we take to be large throughout this study in order to consider rapid yawing. Hence, in this section, we prescribe
\begin{align}
\label{eq: prescribed phi basic case}
\phi(t) = \bar{\phi} + A\sin{\omega t},
\end{align}
where $A$ is the observed amplitude of the angular yawing dynamics and $\bar{\phi}$ is a constant average orientation. Thus, we seek to analyse the system
\begin{subequations}
\label{eq: simplest ODEs prescribed yawing}
\begin{align}
    \diff{x}{t} &= V\cos{\left(\bar{\phi} + A\sin{\omega t}\right)}\,,\label{eq:trans:full_system:x}\\
    \diff{y}{t} &= V\sin{\left(\bar{\phi} + A\sin{\omega t}\right)}\,,\label{eq:trans:full_system:y}
\end{align}
\end{subequations}
where $\omega\gg1$ separates the timescales of oscillation and translation. We will exploit this separation of timescales and pursue a classical multiple-scales analysis \cite{Bender1999}, introducing the fast variable $T\coloneqq \omega t$. In the standard manner, we introduce the formal transformation $t\mapsto(t,T)$ and write $x = x(t,T)$ and $y = y(t,T)$, treating $t$ and $T$ as independent. The additional degree of freedom this affords will be removed later by imposing that our dependent variables are $2\pi$-periodic in $T$. The multiple-scales transformation means that $\mathrm{d}/\mathrm{d}t \mapsto \partial/\partial{}t + \omega\partial/\partial{}T$, which transforms \cref{eq: simplest ODEs prescribed yawing} into the following system of partial differential equations (PDEs):
\begin{subequations}
\label{eq:trans:MS_PDE}
\begin{align}
    \pdiff{x}{t} + \omega \pdiff{x}{T} &= V\cos{\left(\bar{\phi} + A\sin{T}\right)}\,,\label{eq:trans:MS_PDE_x}\\
    \pdiff{y}{t} + \omega \pdiff{y}{T} &= V\sin{\left(\bar{\phi} + A\sin{T}\right)}\,.\label{eq:trans:MS_PDE_y}
\end{align}
\end{subequations}
We proceed by posing asymptotic expansions of our dependent variables of the form
\begin{align}
\label{eq: asymptotic expansions for x and y}
    x = x_0 + \frac{1}{\omega}x_1 + \ldots, \qquad
    y = y_0 + \frac{1}{\omega}y_1 + \ldots \qquad \text{ as }\omega\rightarrow\infty\,,
\end{align}
where the $x_i$, $y_i$ $(i=0, 1, \ldots)$ are ostensibly functions of both $t$ and $T$. Substituting the expansions of \cref{eq: asymptotic expansions for x and y} into \cref{eq:trans:MS_PDE} and comparing terms of $\bigO{\omega}$ gives the leading-order problems
\begin{subequations}
\begin{align}
    \diff{x_0}{T} &= 0\,,\\
    \diff{y_0}{T} &= 0\,,
\end{align}
\end{subequations}
so that $x_0$ and $y_0$ are independent of $T$. Hence, we write $x_0 = \bar{x}_0(t)$ and $y_0 = \bar{y}_0(t)$, adopting the $\bar{\cdot}$ notation to represent functions independent of $T$ throughout. The dependence of these functions on the slow time $t$, and therefore the leading-order behaviour of the swimmer, is not yet known. To determine this, we must proceed to the next order, with the corresponding PDEs being
\begin{subequations}
\label{eq: second order prescribed yawing}
\begin{align}
    \diff{\bar{x}_0}{t} + \pdiff{x_1}{T} &= V\cos{\left(\bar{\phi} + A\sin{T}\right)}\,,\\
    \diff{\bar{y}_0}{t} + \pdiff{y_1}{T} &= V\sin{\left(\bar{\phi} + A\sin{T}\right)}\,.
\end{align}
\end{subequations}
To obtain appropriate governing equations for $\bar{x}_0$ and $\bar{y}_0$, we integrate these PDEs over $T \in (0, 2\pi)$ and impose periodicity in $T$, which causes the terms involving $x_1$ and $y_1$ to vanish. The subsequent averaged equations are
\begin{subequations}
\label{eq:trans:averaged_ode}
\begin{align}
    \diff{\bar{x}_0}{t} &= V\avg{\cos{\left(\bar{\phi} + A\sin{T}\right)}}\,, \label{eq:trans:averaged_ode_x}\\
    \diff{\bar{y}_0}{t} &= V\avg{\sin{\left(\bar{\phi} + A\sin{T}\right)}}\,,\label{eq:trans:averaged_ode_y}
\end{align}
\end{subequations}
where we have defined the averaging operator over the fast dynamics as
\begin{equation}
    \avg{h} = \frac{1}{2\pi}\int\limits_0^{2\pi}h(t,T)\intd{T},
\end{equation}
for functions $h(t,T)$. We note that the PDE multiple-scales formulation has been transformed back into ODEs for the leading-order solutions in \cref{eq:trans:averaged_ode}, as we would expect. It is instructive to evaluate the averages in \cref{eq:trans:averaged_ode} explicitly. In order to do this, we note the following exact integral results
\begin{equation}\label{eq:trans:avgs_of_sine}
    \frac{1}{2\pi}\int\limits_0^{2\pi}\cos{\left(A\sin{T}\right)}\intd{T} = J_0(A),
    \qquad
    \frac{1}{2\pi}\int\limits_0^{2\pi}\sin{\left(A\sin{T}\right)}\intd{T} = 0\,,
\end{equation}
where $J_0$ is the zeroth-order Bessel function of the first kind. The former follows from standard integral representations of the Bessel function, and the latter follows from antisymmetry. Exploiting the linearity of the averaging operator, the averages on the right-hand side of \eqref{eq:trans:averaged_ode} can be expanded as
\begin{subequations}
\label{eq:trans:expanded_avgs_sine}
\begin{align}
     \avg{\cos{\left(\bar{\phi} + A\sin{T}\right)}} &= \cos{\bar{\phi}}\avg{\cos{\left(A\sin{T}\right)}} - \sin{\bar{\phi}}\avg{\sin{\left(A\sin{T}\right)}}\,, \label{eq:trans:expanded_avgs_sine:cos}\\  
     \avg{\sin{\left(\bar{\phi} + A\sin{T}\right)}} &= \sin{\bar{\phi}}\avg{\cos{\left(A\sin{T}\right)}} + \cos{\bar{\phi}}\avg{\sin{\left(A\sin{T}\right)}}\,.\label{eq:trans:expanded_avgs_sine:sin}
\end{align}
\end{subequations}
Making use of the integral results of \cref{eq:trans:avgs_of_sine} in \cref{eq:trans:expanded_avgs_sine}, we may evaluate the right-hand side of \cref{eq:trans:averaged_ode} to obtain the following ODEs for the leading-order translation:
\begin{subequations}
\label{eq: final results for prescribed sinusoidal yawing}
\begin{align}
    \diff{\bar{x}_0}{t} &= \eff{V} \cos{\bar{\phi}}\,, \\
    \diff{\bar{y}_0}{t} &= \eff{V} \sin{\bar{\phi}}\,,
\end{align}
\end{subequations}
where we define the effective translation speed of the yawing swimmer as $\eff{V} = J_0(A) V$. Comparing these equations with the original system of \cref{eq: simplest ODEs}, we see that the leading-order solution to the yawing system of \cref{eq: simplest ODEs prescribed yawing} is given by the solution to ODEs that are analogous to the non-yawing problem. In particular, the effects of yawing manifest as an effective heading of $\bar{\phi}$ and an effective translation speed of $\eff{V}$, at least to leading order. Notably, these results are independent of the yawing rate $\omega$ when $\omega \gg 1$. We will continue to observe this independence of the yawing rate on the leading-order translation in the more-general problems that we consider later.

\begin{figure}
    \centering
    \begin{overpic}[width=0.9\textwidth]{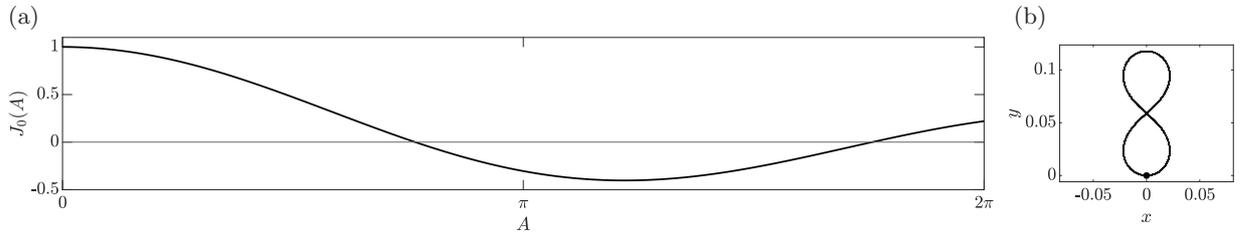}
    \put(0,17){(a)}
    \put(82,17){(b)}
    \end{overpic}
    \caption{A Bessel function and a closed swimming trajectory. (a) The zeroth-order Bessel function of the first kind, $J_0(A)$, plotted as a function of the yawing amplitude $A$, pertinent to the case of sinusoidal yawing. This oscillatory function has magnitude unity only if $A=0$, so that yawing serves to strictly reduce the effective speed of the swimmer, recalling $\eff{V}=J_0(A)V$. Noting the existence of the zeros of $J_0$, we remark that particular values of $A$ will result in a reversed swimming direction or even zero effective speed, though these correspond to very large amplitude yawing, as might be expected. (b) For $A \approx 2.40$, which approximately satisfies $J_0(A)=0$, we simulate the full system of ODEs numerically as described in \cref{sec:trans:general}, realising the closed trajectory predicted by the asymptotic analysis, with displacements on the order of $1/\omega$. Here, $V=1$, $\omega = 20$, and $G(T)=A\sin{T}$, and motion is simulated until $t=10$ from $(x,y)=(0,0)$.}
    \label{fig:bessel_eight}
\end{figure}

Analysing the implications of our results in more detail, since $\abs{J_0(A)} \leqslant 1$, with equality only for $A = 0$, the effect of sinusoidal yawing is to \emph{reduce} the effective translation speed to $\eff{V} = J_0(A) V$, with this reduction being in agreement with intuition. Given the oscillatory nature of the Bessel function, as illustrated in \cref{fig:bessel_eight}a, there are amplitude ranges that would result in a net \emph{reversal} of the translational motion, though these amplitudes may be larger than practically observed in typical microswimmers, requiring $\abs{A}$ greater than approximately 2.40. Moreover, as this Bessel function has non-trivial zeros, a notable implication is that there are specific yawing amplitudes that would result in asymptotically small net motion. In these cases, the rapidly oscillating swimmer trajectory approximately resembles a figure of eight, as shown in \cref{fig:bessel_eight}b, achieving very little net motion. An additional feature of this simple sinusoidal yawing is that the effective heading is precisely $\bar{\phi}$, the average of the prescribed $\phi$ over a single yawing oscillation, though we show below in \cref{sec:trans:general} that this result need not hold for more-general, non-sinusoidal yawing. 

\subsection{General yawing}\label{sec:trans:general}

Though our multiple-scales analysis in the previous subsection was simplified by the choice of sinusoidal yawing, it is straightforward to generalise. Here, we briefly outline the analysis and its implications for the general case.

We consider a general \emph{observed angular yawing} $G(T)$ in place of $A \sin T$ above, and a slowly varying `average' orientation $\bar{\phi}(t)$. These generalisations replace the prescribed $\phi$ in \cref{eq: prescribed phi basic case} with
\begin{equation}
    \phi(t,T) = \bar{\phi}(t) + G(T)\,.
\end{equation}
Our only requirements for $G$ are that it is periodic (we assume it has period $2\pi$, without loss of generality), and that $\avg{G}=0$.

The analysis for this general case proceeds in the same manner as for \cref{sec:trans:sine}, with $A \sin T$ replaced by $G(T)$. Therefore, the $\bigO{1}$ solvability condition is the equivalent of \cref{eq:trans:averaged_ode} and the key integral results that we require are the averages of $\cos{G}$ and $\sin{G}$. These may be succinctly encoded as the real and imaginary parts of the constant complex number
\begin{align}
\label{eq:trans_z}
\trans{z} \coloneqq \avg{\exp{iG}} = \trans{R}\exp{i\trans{\theta}} \in \mathbb{C},
\end{align}
with $R_p$ being the non-negative modulus of $\trans{z}$. This notation allows us to rewrite the drift ODEs for the general case in the compact form
\begin{subequations}
\begin{align}
    \diff{\bar{x}_0}{t} = \eff{V} \cos{\left(\bar{\phi} + \trans{\theta}\right)}\,, \quad
    \diff{\bar{y}_0}{t} = \eff{V} \sin{\left(\bar{\phi} + \trans{\theta}\right)}\,,
\end{align}
where
\begin{align}
\eff{V} \coloneqq \trans{R}V,
\end{align}
\end{subequations}
mirroring the unperturbed swimmer system. Hence, we may define the general effective speed of propulsion as $\eff{V}=\trans{R} V$, and the complex argument $\trans{\theta}$ as the directional bias introduced by the rapid yawing.

The effective speed of propulsion $\eff{V}$ is directly comparable with the specific case of sinusoidal yawing we considered in \cref{sec:trans:sine}, where $\trans{R}\leqslant1$ is a more general version of the specific Bessel function result for sinusoidal yawing, and inherits the accompanying intuition and remarks. Since $\bar{\phi}(t)$ represents the heading of the swimmer in the absence of yaw, we may interpret the complex argument $\trans{\theta}$ as the directional bias introduced by the rapid yawing. This general bias was not present in the case of sinusoidal yaw (with the exception of the potential reversal of the effective swimmer direction). The reason for this is that the directional bias will vanish if $\avg{\sin{G}}=0$, which is the case for $G(T) = \sin(T)$. Intuitively, this condition is equivalent to requiring that the average of the component of the velocity perpendicular to the mean heading is zero, an interpretation that is most easily seen if we consider $\bar{\phi}=0$, wherein $V\sin{G}$ is precisely the $Y$-component of the motion. Similarly, $\avg{\cos{G}}$ quantifies the average component of the velocity in the direction of the average heading, along the $X$-axis if $\bar{\phi}=0$, and, hence, modifies the swimming speed.

\begin{figure}
    \centering
    \begin{overpic}[width=0.7\textwidth]{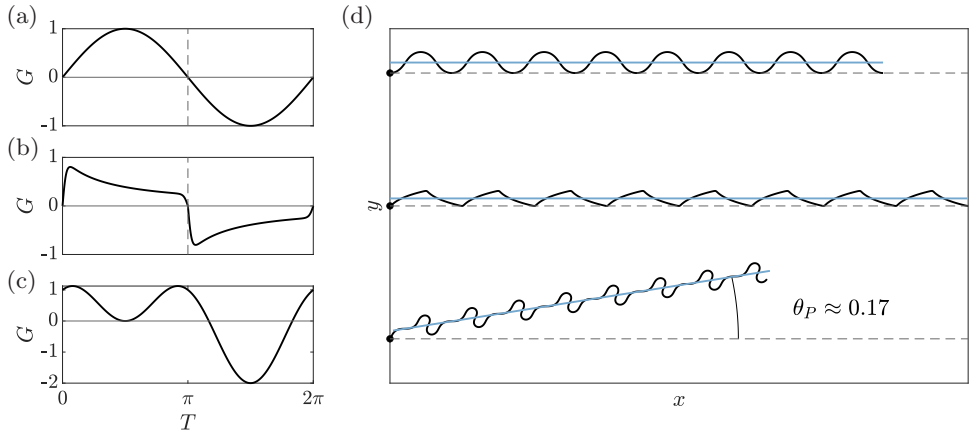}
    \put(-1,43){(a)}
    \put(-1,29){(b)}
    \put(-1,15){(c)}
    \put(34,43){(d)}
    \end{overpic}
    \caption{Exemplar yawing functions $G(T)$ and the associated translational motion. (a) A sinusoidal yawing function, which is odd about $T_o^{\star}=\pi$, shown dashed. (b) An antiperiodic yawing function, satisfying $G(T+\pi)=-G(T)$, with $G$ on $[0,\pi]$ being the reflection of $G$ on $[\pi,2\pi]$ along the $T$ axis. (c) A biased yawing function, satisfying neither of the presented antisymmetry conditions and with $\avg{\sin{G}}\neq0$. (d) The numerical solutions of the full yawing problem, taking $\phi = G(T)$ so that the direction of motion in the absence of yaw is along the $X$-axis, shown for each of the considered yawing functions and offset vertically for visual clarity. Having simulated the motion in each case for the same amount of time, the dependence of the effective speed on the form of the yawing function is apparent, with case (b) giving rise to the largest $\eff{V}$ and, hence, the greatest displacement from the initial configuration. Cases (a) and (b), which satisfy $\avg{\sin{G}}=0$, give rise to unbiased motion, whilst the yawing function of panel (c) yields biased motion that is not aligned along the axis, despite the average swimmer orientation $\phi$ being zero. In blue, we plot the leading-order solution of the multiple-scales analysis in each case, demonstrating excellent agreement with the full yawing behaviour, and have made use of the asymptotically corrected initial condition of \cref{sec:drift}. All considered yawing functions satisfy $\avg{G}=0$ and we have fixed $\omega=40\pi$ for simulation.}
    \label{fig:G_examples}
\end{figure}

It is interesting to consider forms of generalised yawing that lead to unbiased swimming, through our derived condition $\avg{\sin{G}}=0$. In particular, we may identify two general classes of antisymmetric dynamics that give rise to unbiased swimming. The first category is simply that of all odd functions, satisfying $G(T_o^{\star}+T)=-G(T_o^{\star}-T)$ for some $T_o^{\star}$, which includes $G(T) = A\sin{T}$ with $T_o^{\star}=0$, for instance. The second category is all functions that satisfy $G(T+\pi) = -G(T)$, also known as \emph{antiperiodic} functions, a condition that captures the notion of mirror symmetry of the yawing over a period. In \cref{fig:G_examples}, we present exemplar $G(T)$ that satisfy each of these conditions, and note that, for any given $G(T)$, the quantity $\trans{z}$ can be easily computed numerically, yielding the effective propulsive velocity and the directional bias. Here and throughout, systems of ordinary differential equations are solved via \texttt{ode45}, a standard routine in MATLAB\textsuperscript{\textregistered}\cite{Shampine1997}. Further, for any swimmer whose geometrical configuration in a swimmer-fixed reference frame at time $T+\pi$ is simply a reflection of the swimmer configuration at time $T$ in a swimmer-fixed plane, such as the model spermatozoon showcased in \cref{app:sperm_validity}, we can define the swimmer orientation such that the yawing function $G(T)$ will be antiperiodic. This follows from symmetry and the time-independence of Stokes equations, as described in more detail in \cref{app:antiperiodic swimmers}. Hence, we may immediately conclude that the yawing of such swimmers does not contribute any directional bias to their motion at leading order, at least in the context of this simplified model of motion.

As is common in the method of multiple scales, it is pertinent to understand the effects of longer-term drift of the full solution away from the derived leading-order expansion. We investigate this numerically (not shown), finding that the leading-order solution remains accurate on significant timescales, over tens of thousands of yawing oscillations, explored for $\omega=40\pi$, $V=1$, and $A=1$ in the case of sinusoidal yawing. Given that we have only calculated solutions up to leading order, we would expect an accumulating error at $\bigO{1/\omega}$. However, we do not observe this. That is, our asymptotic analysis appears to be more accurate than we might expect it to be. We will later return to this point in \cref{sec:drift} and in \cref{app:3D_trans}, in which we justify this surprising improved accuracy by consideration of the higher-order problem.

Having set up and illustrated the general framework for considering rapid yawing motions, we next extend our analysis to a class of bodies in the presence of a canonical background flow. This will mean that the angular motion of the swimmer, $\phi$, is \emph{coupled} to the external flow, such we will no longer be able to directly prescribe the yawing rotational dynamics, as we did above.

\section{Yawing in shear flow}\label{sec:rot}
Described in the seminal work of \citeyear{Jeffery1922}, Jeffery's orbits capture the motion of a rigid torque-free ellipsoid in a background shear flow with a no-slip condition on its surface \cite{Jeffery1922}. Through the solution of the governing Stokes equations, \citeauthor{Jeffery1922} derived ordinary differential equations in terms of an Euler angle formulation, and in some cases noted their exact solution. The relevant nonlinear equation for the evolution of the azimuthal angle $\phi$ decouples from the other Euler angles, and is
\begin{equation}\label{eq:Jeffery}
    \diff{\phi}{t} = \dfrac{\gamma}{2}\left(1 - B \cos{2\phi}\right) := f(\phi;B)\,.
\end{equation}
Here, $\gamma$ denotes the constant shear rate of the background flow, $B$ is a constant derived from the geometrical properties of the ellipsoid, and we will impose an initial condition of $\phi(0) = 0$ for notational simplicity, though this may be readily substituted. The geometrical parameter $B$ differs from the notation of \citet{Jeffery1922}, instead reflecting the significant generalisation of Jeffery's orbits by \citet{Bretherton1962}, which extended the classical description to include bodies of revolution, amongst additional contributions. As such, we will refer to $B$ as the Bretherton constant, which may assume any real value. For $\abs{B}<1$, the dynamics of \cref{eq:Jeffery} are precisely those of the classic Jeffery's orbits, with periodic rotation occurring ad infinitum; for $\abs{B}>1$, the solution approaches one of two stable steady states, as discussed by \citet{Bretherton1962} and more recently by \citet{Ishimoto2020b}. This latter work, and its recent generalisation \cite{Ishimoto2020a}, further extended the Jeffery's orbit description to bodies with particular symmetries, far removed from the original ellipsoidal considerations of Jeffery. Despite this, the solution to \cref{eq:Jeffery} can be written in the simple form found by Jeffery, which we write as
\begin{equation}\label{eq:Jeffery_solution}
    \phi(t) = 
    \phi_J(t;B) = \arctan{\left(\frac{1}{r}\tan{\frac{\gamma t}{r + \frac{1}{r}}}\right)}\,, \quad r \coloneqq \sqrt{\frac{1+B}{1-B}}\in\mathbb{C}\,,
\end{equation}
where $r$ is precisely the aspect ratio of the rigid body if the object in question is a spheroid. From \cref{eq:Jeffery_solution}, the distinction between the cases of $\abs{B}<1$ and $\abs{B}>1$ is clear, corresponding to $r$ being real or imaginary, respectively. With reference to its historical origin, we will refer to this solution as a Jeffery's orbit irrespective of the magnitude of $B$, with the understanding that this `orbit' may in fact lead to a steady solution for bodies with Bretherton constants of magnitude larger than unity.

In the case of planar motion, this solution completely describes the angular evolution of the rigid torque-free object in flow.  Owing to the linearity of Stokes equations, we may readily include consideration of these angular dynamics into the yawing self-propelled swimmer system of the previous section, leading to the nonlinear dynamical system
\begin{subequations}
\label{eq:full_system}
\begin{align}
    \diff{x}{t} &= V \cos{\phi} - \gamma y\,,\label{eq:full_system_x}\\
    \diff{y}{t} &= V \sin{\phi}\,,\label{eq:full_system_y}\\
    \diff{\phi}{t} &= \frac{\gamma}{2}\left(1 - B \cos{2\phi}\right) + \omega G'(\omega t)\,.\label{eq:full_system_phi}
\end{align}
\end{subequations}
Here, we have included the effects of a background shear flow, with velocity $\vec{u}=-\gamma Y\vec{e}_X$ at a point $(X,Y)$ in the plane, driving the rotation and contributing to the translation. We emphasise that the motion of a passive object in a shear flow can be obtained by taking $V=0$ in what follows. Similarly, the results of the previous section without a background flow can be recovered by taking $\gamma = 0$ in what follows. In full generality, the swimming direction need not coincide with a context-dependent definition of the swimmer orientation, though this is assumed in \cref{eq:full_system} to afford notational simplicity. Consistent with this, in what follows, we will assume that the orientation and direction of the swimmer are both captured by $\phi$, noting that a constant difference between these quantities can be trivially included in the dynamics, mapping $\phi\mapsto\phi + C$ for constant discrepancy $C$ in any equations of rotational motion, for instance, with no significant consequence.

The general yawing dynamics of \cref{sec:trans:general} are included in \cref{eq:full_system_phi} via the derivative of the yawing function, denoted by $G'$. As above, we impose vanishing initial conditions for convenience, but note that these may be readily generalised. Without the yawing, the augmented dynamical system of \cref{eq:full_system} represents a minimal model of an active particle in a background shear flow, and has previously been used to model the average motion of complex, flagellated swimmers. One such example is in the context of the monoflagellated parasite \leishmex{} \cite{Walker2018}, where a similar model was used to investigate the long-time behaviours of a cell swimming near a boundary. Further, in \cref{app:sperm_validity}, we evidence the numerical validity of such a minimal model when seeking to study human spermatozoa by comparison with a direct numerical simulation of the angular microswimmer dynamics. We do note, however, that a rigorous theoretical exploration of the surprising general efficacy of such reduced representations remains an interesting avenue for future work. Acknowledging this, we henceforth assume the applicability of \cref{eq:Jeffery} in modelling the angular dynamics of a yaw-free microswimmer in a background shear flow.

In a similar manner to the previous section, we seek a multiple-scales solution to the yawing system of \cref{eq:full_system}. Under the same transformation, \cref{eq:full_system} becomes
\begin{subequations}
\label{eq:full_transformed_system}
\begin{align}
    \pdiff{x}{t} + \omega \pdiff{x}{T} &= V \cos{\phi} - \gamma y\,,\label{eq:full_transformed_system_x}\\
    \pdiff{y}{t} + \omega \pdiff{y}{T} &= V \sin{\phi}\,,\label{eq:full_transformed_system_y}\\
    \pdiff{\phi}{t} + \omega \pdiff{\phi}{T} &= \frac{\gamma}{2}\left(1 - B \cos{2\phi}\right) + \omega G'(T)\,.\label{eq:full_transformed_system_phi}
\end{align}
\end{subequations}

Here, noting the decoupling of the angular evolution equation from those of translation, we will focus on presenting an analysis of the angular dynamics, with the leading-order translational motion being constructed analogously to \cref{sec:trans:general} with only minimal modification. We pose an asymptotic expansion for $\phi$ of the form
\begin{equation}
    \phi = \phi_0 + \frac{1}{\omega}\phi_1 + \cdots \quad \text{ as } \omega\rightarrow\infty\,,
\end{equation}
and equate coefficients of $\omega$ in \cref{eq:full_transformed_system_phi}. The leading-order angular problem is simply
\begin{equation}
    \pdiff{\phi_0}{T} = G'(T)\,,
\end{equation}
which has general solution
\begin{equation}\label{eq: phi_0 general solution}
    \phi_0 = \bar{\phi}_0(t) + G(T)\,,
\end{equation}
precisely mirroring the form of $\phi$ considered in \cref{sec:trans:general}. However, the key difference is that the slow dynamics, $\bar{\phi}_0(t)$, are currently \emph{undetermined} rather than being prescribed, necessitating an analysis of the higher-order problem to determine the angular evolution.

At the next order in $\omega$, the appropriate terms in \cref{eq:full_transformed_system_phi} are
\begin{equation}
\label{eq:second-order-angular-ODE}
    \diff{\bar{\phi}_0}{t} + \pdiff{\phi_1}{T} = f(\phi_0;B)\,,
\end{equation}
recalling our shorthand $f(\phi;B)$ for the forcing of Jeffery's ODE \cref{eq:Jeffery} in the absence of yaw. Averaging \cref{eq:second-order-angular-ODE} over a period of yawing and imposing periodicity in $T$, we arrive at the averaged ordinary differential equation
\begin{equation}
\label{eq:leading_order_ODE_phi0_unchanged}
    \diff{\bar{\phi}_0}{t} = \avg{f(\phi_0;B)} = \frac{\gamma}{2}\left(1 - B\avg{\cos{2\phi_0}}\right)\,,
\end{equation}
which governs $\bar{\phi}_0$. We encountered the average of a slightly different cosine term in \cref{sec:trans:general}; following a similar analysis allows us to encode the key integral results that we require as the real and imaginary parts of the complex number
\begin{align}
\label{eq:rot_z}
\rot{z}\coloneqq\avg{\exp{2iG}}=\rot{R}\exp{i\rot{\theta}} \in \mathbb{C},
\end{align}
similar to the constant $\trans{z}$ from \cref{eq:trans_z}. Such an analysis allows us to re-write \cref{eq:leading_order_ODE_phi0_unchanged} in terms of $\rot{z}$ as
\begin{subequations}
\label{eq:leading_order_ODE_phi0}
\begin{equation}
    \diff{\bar{\phi}_0}{t} = \frac{\gamma}{2}\left(1 - \eff{B}\cos{\left(2\bar{\phi}_0 + \rot{\theta}\right)}\right) = f(\bar{\phi}_0 + \rot{\theta}/2;\eff{B})\,,
\end{equation}
where
\begin{align}
\eff{B}\coloneqq\rot{R}B
\end{align}
\end{subequations}
is the effective Bretherton constant and $\rot{\theta}/2$ is an effective phase shift. Comparing \cref{eq:leading_order_ODE_phi0} to that derived by Jeffery, recounted here in \cref{eq:Jeffery}, we see that this leading-order solution corresponds to the Jeffery's orbit of a biased angular variable $\psi\coloneqq\bar{\phi}_0 + \rot{\theta}/2$, i.e.
\begin{equation}\label{eq: redefined orientation}
    \diff{\psi}{t} = f(\psi;\eff{B})\,.
\end{equation}
Therefore, we have shown that any prescribed yawing function $G$ within a Jeffery's orbit will lead, on average, to Jeffery's orbit dynamics at leading order with an effective Bretherton constant $\eff{B}$, where $\abs{\eff{B}}\leqslant\abs{B}$, and with a phase shift of $\rot{\theta}/2$, which is non-zero if $\avg{\sin{2G}}\neq0$.

This has particular implications given the markedly different character of Jeffery's orbits associated with $\abs{B}>1$ and $\abs{B}<1$. Our results suggest that the introduction of yaw can transition an object from one regime to another. In particular, since yaw strictly reduces the effective magnitude of the Bretherton constant, yawing can elicit non-trivial periodic orbits from bodies that may otherwise have approached a steady state in the background flow. Such a transition is dependent only on the original Bretherton constant and the details of the yawing gait, and is independent of the yawing frequency at leading order. In special cases, for gaits that satisfy $\avg{\sin{2G}}=0$, there is no phase shift and we recover standard Jeffery's orbit dynamics, with the Bretherton constant being scaled by $\avg{\cos{2G}}$. With reference to the discussion of \cref{sec:trans:general}, such a condition is automatically satisfied by yawing functions that satisfy either $G(T_o^{\star}+T)=-G(T_o^{\star}-T)$ or $G(T+\pi) = -G(T)$, for some fixed $T_o^{\star}$.

In summary, the leading-order system of ODEs for the translational and angular yawing problem in the presence of a shear flow is given by
\begin{subequations}\label{eq: leading_order drift ODEs}
\begin{align}
    \diff{\bar{x}_0}{t} &= \eff{V}\cos{\left(\bar{\phi}_0 + \trans{\theta}\right)} - \gamma \bar{y}_0\,,\\
    \diff{\bar{y}_0}{t} &= \eff{V}\sin{\left(\bar{\phi}_0 + \trans{\theta}\right)}\,,\\
    \diff{\bar{\phi}_0}{t} &= \frac{\gamma}{2}\left(1 - \eff{B}\cos{\left(2\bar{\phi}_0 + \rot{\theta}\right)}\right)\,,
\end{align}
\end{subequations}
identical to the dynamics in the absence of yawing except for the introduction of: (1) an effective propulsive speed, $\eff{V}$; (2) a bias to the direction of propulsion, $\trans{\theta}$; (3) an effective geometry via the effective Bretherton constant, $\eff{B}$; (4) a bias in the Jeffery's orbit dynamics, $\rot{\theta}/2$. The angular biases for the translational and rotational dynamics vanish if the yawing gait satisfies $\avg{\sin{G}}= 0$ and $\avg{\sin{2G}}=0$, respectively. We note that these conditions are automatically satisfied for any antiperiodic $G$.

For the sinusoidal case, taking $G(T) = A\sin{T}$, it is straightforward to compute the following effective coefficients: $\eff{V} = J_0(A)V$, $\trans{\theta} = 0$, $\eff{B}=J_0(2A)B$, and $\rot{\theta} = 0$. Thus, the corresponding leading-order system is given by
\begin{subequations}
\begin{align}
    \diff{\bar{x}_0}{t} &= \eff{V}\cos{\phi_J(t;\eff{B})} - \gamma \bar{y}_0\,,\\
    \diff{\bar{y}_0}{t} &= \eff{V}\sin{\phi_J(t;\eff{B})}\,,\\
    \phi_0(t) &= \phi_J(t;\eff{B}) + A\sin{\omega t}\,, \label{eq:leading_order_angular_solution}
\end{align}
\end{subequations}
where $\phi_0$ corresponds to a Jeffery's orbit with modified Bretherton constant and additive high-frequency yawing.

\begin{figure}
    \centering
    \begin{overpic}[percent, width = 0.9\textwidth]{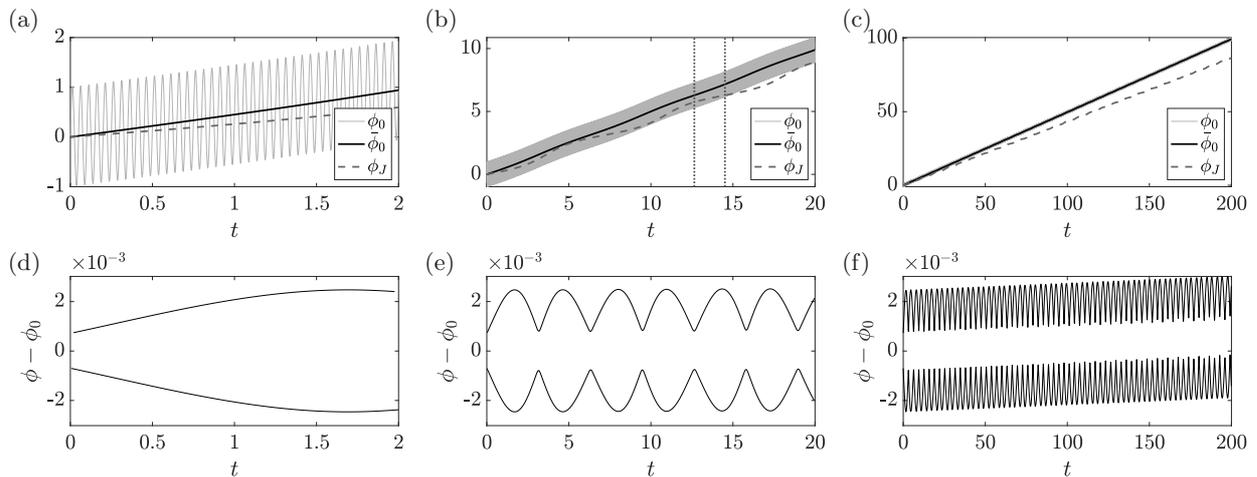}
       \put(-1,37){(a)}
       \put(33,37){(b)}
       \put(67,37){(c)}
       \put(-1,17){(d)}
       \put(33,17){(e)}
       \put(67,17){(f)}
    \end{overpic}
    \caption{The angular evolution of a yawing body. (a,b,c) The leading-order multiple-scales solution $\phi_0$ is shown over time alongside the leading-order solution averaged over a single oscillation $\bar{\phi}_0$, though is not visible at the resolution of (c). Also plotted is the solution to the unperturbed dynamics, $\phi_J$, from which we observe significant differences between the yawing and non-yawing systems. The periods of the slow dynamics are shown as dotted vertical lines in (b), with the yawing giving rise to a shorter period since $\abs{\eff{B}}<\abs{B}$. (d,e,f) The envelope of the difference between the numerical solution to the full perturbed evolution equation, denoted by $\phi$, and the leading-order multiple-scales solution $\phi_0$. We note that this error is on the order of $10^{-3}$ across the timescales of each of these plots, validating the accuracy of the leading-order multiple-scales solution, with the accumulation of error over time only noticeable in the very longest comparison, which corresponds to approximately 16 full revolutions in the background flow, suggesting a long-term efficacy of the leading-order solution. Here, we have fixed $\gamma=1$, $B=0.5$, $A=1$, and $\omega=40\pi$.}
    \label{fig:angular_verif}
\end{figure}

We compare our asymptotic results to numerical solutions of the full system of \cref{eq:full_system}, adopting sinusoidal yawing. We fix the parameters to be $\gamma=1$, $B=0.5$, $A=1$, and $\omega=40\pi$, the latter corresponding to a frequency of 20 oscillations per unit time. In \cref{fig:angular_verif}a-c, which display the dynamics on different timescales, respectively, we plot both the leading-order solution $\phi_0$, given by \cref{eq:leading_order_angular_solution}, and the averaged solution $\bar{\phi}_0 = \phi_J(t;\eff{B})$. For comparison, we also plot the solution to the non-yawing problem, which is simply $\phi_J(t;B)$, i.e. a Jeffery's orbit with differing Bretherton constant, with \cref{fig:angular_verif}b illustrating the significant effects that rapid yawing can have on the dynamics over even a single long-time rotation in the background shear flow, including the period of rotation. At the resolution of these plots, the leading-order solution and the numerical solution of the yawing problem \cref{eq:full_system_phi} would be indistinguishable, a property captured by \cref{fig:angular_verif}d-f, which display the error $\phi - \phi_0$. We systematically evaluate this error for various $\omega$ in \cref{app:error_scaling}, validating the asymptotic analysis and observing the expected $\bigO{1/\omega}$ scaling. For the parameters of the exploration of \cref{fig:angular_verif}, the error in the leading-order approximation is on the order of $10^{-3}$, a feature that is retained even over the longest timescale, which constitutes approximately 16 full revolutions in the background flow and 4000 yawing oscillations in this case. With the persistence of this low-magnitude error being somewhat unusual, we now move to investigate the higher-order problem, seeking to understand the source of this prolonged accuracy.


\section{Higher-order drift and bias}\label{sec:drift}

In the previous sections, we constructed leading-order problems for the translational and rotational motion of a yawing swimmer. When comparing our leading-order asymptotic solutions to numerical results of the full dynamics, we found that our asymptotic solutions were significantly more accurate than one might expect. In order to explore this observation, we now seek to calculate the slow drift in the system at $\bigO{1/\omega}$, the next asymptotic order in the system.

We consider the most general system of \cref{eq:full_transformed_system}, with leading-order solutions constructed in \cref{sec:rot}. To calculate the higher-order solutions, we begin with the $\bigO{1}$ terms in \cref{eq:full_transformed_system}, which we recount as
\begin{subequations}
\label{eq:full_transformed_system O 1/w}
\begin{align}
     \pdiff{x_1}{T} + \pdiff{x_0}{t} &= V \cos{\phi_0} - \gamma y_0\,,\label{eq:full_transformed_system_x O 1/w}\\
    \pdiff{y_1}{T} + \pdiff{y_0}{t} &= V \sin{\phi_0}\,,\label{eq:full_transformed_system_y O 1/w}\\
    \pdiff{\phi_1}{T} + \pdiff{\phi_0}{t} &= \frac{\gamma}{2}\left(1 - B \cos{2\phi_0}\right)\,.\label{eq:full_transformed_system_phi O 1/w}
\end{align}
\end{subequations}
Previously, we integrated these equations over a period in $T$ to yield leading-order solvability conditions and the corresponding solutions for $x_0$, $y_0$, and $\phi_0$. In order to proceed to higher order, we instead solve these equations to obtain
\begin{subequations}
\label{eq: phi_1 and I phi solution}
\begin{align}
\label{eq: phi_1 solution}
x_1(t,T) = \Ix(t,T) - \avg{\Ix} + \bar{x}_1(t), \\
y_1(t,T) = \Iy(t,T) - \avg{\Iy} + \bar{y}_1(t), \\
\phi_1(t,T) = \Iphi(t,T) - \avg{\Iphi} + \bar{\phi}_1(t),
\end{align}
\end{subequations}
where
\begin{subequations}
\label{eq: I definitions}
\begin{align}
\Ix(t,T) &\coloneqq V \int\limits_0^{T} \! \left[\cos \phi_0 - \avg{\cos \phi_0}\right] \intd{s} - \gamma \int\limits_0^{T} \! \left[y_0 - \avg{y_0}\right] \intd{s}\,, \\
\Iy(t,T) &\coloneqq V \int\limits_0^{T} \! \left[\sin \phi_0 - \avg{\sin \phi_0}\right] \intd{s}\,, \\
\Iphi(t,T) &\coloneqq \dfrac{\gamma B}{2} \int\limits_0^T \! \left[\avg{\cos 2 \phi_0} - \cos 2 \phi_0 \right] \intd{s}\,.
\end{align}
\end{subequations}
Here, $\bar{x}_1$, $\bar{y}_1$, and $\bar{\phi}_1$ represent the second-order drifts, and are all as-of-yet undetermined functions of the slow time. 

There are several ways to determine these higher-order drifts. A standard method would be to repeat the approach employed earlier, collecting the $\bigO{1/\omega}$ terms in \cref{eq:full_transformed_system} and deriving solvability conditions by integrating over $T$ and imposing periodicity. An alternative approach, applicable to this particular problem and affording significant notational simplicity over the previous method, is to derive ODEs for the combined drift, at both leading and sub-leading orders, from the outset. To do this, we integrate the full dynamical system of \cref{eq:full_transformed_system} over $T$, before considering any particular asymptotic expansions, and impose $T$-periodicity, obtaining
\begin{subequations}
\label{eq:full_transformed_system averaged}
\begin{align}
     \diff{\avg{x}}{t} &= V \avg{\cos{\phi}} - \gamma \avg{y}\,,\label{eq:full_transformed_system_x averaged}\\
     \diff{\avg{y}}{t} &= V \avg{\sin{\phi}}\,,\label{eq:full_transformed_system_y averaged}\\
     \diff{\avg{\phi}}{t} &= \frac{\gamma}{2}\left(1 - B \avg{\cos{2\phi}}\right)\,.\label{eq:full_transformed_system_phi averaged}
\end{align}
\end{subequations}
These differential equations will govern the evolution of the full average dynamics, into which we can substitute the previously identified asymptotic expansions for $x$, $y$, and $\phi$. In particular, defining the convenient notation
\begin{align}
\label{eq: slow time drift functions}
\bar{x}(t) = \bar{x}_0(t) + \dfrac{1}{\omega}\bar{x}_1(t), \quad
\bar{y}(t) = \bar{y}_0(t) + \dfrac{1}{\omega}\bar{y}_1(t), \quad
\bar{\phi}(t) = \bar{\phi}_0(t) + \dfrac{1}{\omega}\bar{\phi}_1(t),
\end{align}
for the slowly-evolving drift of the dynamics, recalling the asymptotic solutions of the previous section, we write
\begin{align}
\label{eq: redefined asymptotic expansions}
x \sim \bar{x}(t) + \dfrac{1}{\omega}\left( \Ix(t,T) - \avg{\Ix}\right), \quad
y \sim \bar{y}(t) + \dfrac{1}{\omega}\left( \Iy(t,T) - \avg{\Iy}\right), \quad
\phi \sim \bar{\phi}(t) + G(T) + \dfrac{1}{\omega} \left( \Iphi(t,T) - \avg{\Iphi}\right),
\end{align}
with asymptotic errors of $\bigO{1/\omega^2}$. Inserting these asymptotic expansions into the differential equations governing the average dynamics, \cref{eq:full_transformed_system averaged}, and retaining terms of both $\bigO{1}$ and $\bigO{1/\omega}$, we find
\begin{subequations}
\label{eq:joint drift}
\begin{align}
     \diff{\bar{x}}{t} &= V \trans{\tilde{R}} \cos\left(\bar{\phi} + \trans{\tilde{\theta}}\right) - \gamma \bar{y}\,,\label{eq:joint drift x}\\
     \diff{\bar{y}}{t} &= V \trans{\tilde{R}} \sin\left(\bar{\phi} + \trans{\tilde{\theta}}\right)\,,\label{eq:joint drift y}\\
     \diff{\bar{\phi}}{t} &= \frac{\gamma}{2}\left(1 - B \rot{\tilde{R}} \cos \left(2 \bar{\phi} + \rot{\tilde{\theta}}\right) \right)\,,\label{eq:joint drift phi}
\end{align}
\end{subequations}
where
\begin{subequations}
\label{eq: modified za zp}
\begin{align}
 \trans{\znew}(t) \coloneqq \avg{\exp{\left[i\left\{G(T) + \dfrac{1}{\omega} \left(\Iphi(t,T) - \avg{\Iphi} \right) \right\}\right]}} = \trans{\tilde{R}} \exp{i \trans{\tilde{\theta}}} \in \mathbb{C}, \\
\rot{\znew}(t) \coloneqq \avg{\exp{\left[ 2i\left\{G(T) + \dfrac{1}{\omega} \left(\Iphi(t,T) - \avg{\Iphi} \right) \right\}\right]}} = \rot{\tilde{R}} \exp{i \rot{\tilde{\theta}}} \in \mathbb{C}.
\end{align}
\end{subequations}
Consistent initial conditions for the ODEs of \cref{eq:joint drift} can be determined by combining \cref{eq: redefined asymptotic expansions,eq: slow time drift functions,eq: phi_1 and I phi solution}, yielding
\begin{align}\label{eq: higher order init conds}
\left(\bar{x}(0), \bar{y}(0), \bar{\phi}(0) \right) = \left(x(0), y(0), \phi(0) \right) + \dfrac{1}{\omega} \left.\left(\avg{\Ix}, \avg{\Iy}, \avg{\Iphi} \right)\right|_{t = 0}.
\end{align}

Having arrived at these equations for the evolution of the average dynamics, we are able to conclude that the long-time drift, captured by $\bar{x}$, $\bar{y}$, and $\bar{\phi}$ and including terms of order $\bigO{1/\omega}$, satisfies equations of precisely the same form as the leading-order system of \cref{eq: leading_order drift ODEs}. Hence, the conclusions of the previous section, that yawing modifies the average dynamics by modifying the swimmer shape, reducing the swimming speed, and potentially introducing orientational biases, hold even at $\bigO{1/\omega}$. Moreover, any possible deviations away from this must appear at $\bigO{1/\omega^2}$ or higher. 

Further, we are able to more-precisely quantify the effects of the yawing, with biases and reductions in both swimming and shape parameters now captured at both leading and sub-leading asymptotic orders. It is straightforward to show that $\trans{\znew} \sim \trans{z}$ and $\rot{\znew} \sim \rot{z}$ as $\omega \to 0$, as we would expect. The more interesting question is the form of the next-order terms of these modified coefficients, that is, at $\bigO{1/\omega}$. A detailed analysis of the next-order terms of \cref{eq: modified za zp} is presented in \cref{app:drift_details}, and we now summarise the key conclusions of this calculation.



Firstly, we find that $\rot{\znew} - \rot{z} = \smallO{1/\omega^2}$ if $G$ is antiperiodic. Thus, the average leading-order solution for the angular dynamics $\bar{\phi}_0$ obeys the same ODE as the second-order-accurate solution $\bar{\phi}$, and these solutions are equal subject to imposing the higher-order initial condition of \cref{eq: higher order init conds} on $\bar{\phi}_0$. We further note that even if one were to simply employ leading-order initial conditions, the $\bigO{1/\omega}$ error introduced would be periodic and thus bounded in time, so that any long-term growth in error would still occur at $\bigO{1/\omega^2}$ or higher, in accordance with the numerical explorations shown in \cref{fig:angular_verif}. Additionally, recalling the conclusions of \cref{sec:rot}, the imposed antiperiodicity condition is also sufficient to give rise to unbiased angular dynamics, with $\rot{\theta}=0$ and $\rot{\tilde{\theta}} = \smallO{1/\omega}$.

Secondly, now with reference to the translational dynamics, we find that antiperiodicity alone is not sufficient to yield higher-order agreement between the leading-order and full solutions. However, imposing the additional requirement of $G$ being even about some $T_e^{\star}$ is sufficient to align the average leading-order solution and the higher-order averaged solution, again subject to the prescription of the corrected initial conditions of \cref{eq: higher order init conds}. Indeed, if the initial condition of the rotational dynamics is specified to next order, but the initial condition of the translational dynamics is only given at leading order, then the resulting error in the translational dynamics at $\bigO{1/\omega}$ will only remain constant in time if $\gamma=0$, i.e. if there is no background flow. This is a direct result of the background flow amplifying errors in the approximate solutions, with the $\gamma y$ term of \cref{eq:full_system_x} introducing error that grows with time as the swimmer is advected by the flow.

Finally, in the absence of flow and with a prescribed orientation, such as in the case of pure translation in \cref{sec:trans}, we may readily conclude from \cref{eq:full_transformed_system averaged} that there is no error in the dynamical system at any order, with the averages of $\cos{G}$, $\sin{G}$, and $\cos{2G}$ being known exactly. These conclusions hold at all asymptotic orders and are not reliant on any assumptions on the yawing function, as described in the extended context of three-dimensional motion in \cref{app:3D_trans}. However, due to the potential appearance of additional slower timescales from using the method of multiple scales, we refrain from concluding that the leading-order solution for pure translation, subject to the second-order initial conditions of \cref{eq: higher order init conds}, is accurate to all orders.

\section{Discussion}
In this work, we have theoretically and numerically explored the effects of high-frequency yawing oscillations on both objects in Stokes flow and the motion of active particles, from the dynamics of a self-propelled particle to the canonical Jeffery's orbit. In this latter example, employing the asymptotic method of multiple scales as we have done throughout, we have found, perhaps surprisingly, that the slow angular dynamics in the presence of rapid yawing are once again governed by Jeffery's equation in the plane. This supports, though does not justify, the use of Jeffery's orbits for the modelling of yawing bodies, as introducing oscillations to a non-yawing body undergoing a Jeffery's orbit preserves the structure of the dynamical system, a feature that also holds for the translational dynamics. In particular, the general effect of the yawing on the slow dynamics is to: (1) modify the effective shape of the particle, (2) introduce a phase shift, and (3) strictly reduce the magnitude of the Bretherton constant and any non-zero swimming speed. Notably, the precise frequency of the rapid oscillation does not affect the slow dynamics at leading order. The potential phase shift in the dynamics can be seen to vanish if the yawing is antiperiodic, with the effects of rapid oscillations otherwise motivating an alternative definition of body orientation that is applicable to yawing objects and explicitly accounts for this phase shift.

In the context of spheroids, which have Bretherton constant of magnitude less than unity, we have seen that high-frequency yawing gives rise to an effective shape that is less elongated, with the extent of this change being governed by the quantity $\rot{R}$, defined in \cref{eq:rot_z}. For bodies with Bretherton constant of magnitude greater than one, which can involve fairly unusual shapes (as illustrated by \citet{Bretherton1962}), the effects of yawing can result in significant qualitative differences. In particular, rapid yawing can alter the long-term behaviour of the body, since the solutions of Jeffery's ordinary differential equation are of markedly different character for $\abs{B}<1$ and $\abs{B}>1$; they are non-trivially periodic in the former and approach a constant in the latter \cite{Singh2013, Borker2019}. Hence, rapid oscillations can drive qualitative changes in the dynamics of an object in shear flow, and otherwise serve to reduce their period of oscillation. This investigation into a perturbation of a Jeffery's orbit complements the large body of research that utilises or analyses the motion of particles in flow, with particular reference to the generalising studies of \citet{Bretherton1962} and \citet{Ishimoto2020a}, amongst others \cite{Chen2011, Fries2017, Thorp2019}.

The condition of \emph{antiperiodic} yawing has been a key property several times during this manuscript. We find that antiperiodicity removes orientational biases from translational and angular motion and bounds the dynamics at next order, resulting in a long-time accuracy of the leading-order dynamics derived in this study, a non-standard property in general multiple-scales approaches. As noted at the conclusion of \cref{sec:trans:general}, this condition arises from stereotypical beating patterns of canonical swimmers, such as the amplitude-modulated travelling waves of displacement occurring in spermatozoa, and was indeed found to be satisfied in the computational exploration of \cref{app:sperm_validity} for a model spermatozoon. Hence, whilst not a necessary condition for removing orientational bias or affording additional accuracy to our approximations, this constraint of antiperiodicity appears to feature in numerous swimming gaits. In particular, as discussed in \cref{app:antiperiodic swimmers}, antiperiodicity of yawing arises from bodies with symmetric swimming gaits, though only if the swimmer orientation is defined such that the average of the yawing function vanishes. In turn, this suggests that there is some notion of a natural definition of swimmer orientation, dependent on the geometry of the swimmer and the details of the swimming gait. With a universal concept of swimmer orientation lacking, leading to definitions that vary from one study to the next, seeking to satisfy this constraint outside the context of the models considered in this work may serve to reduce the freedom in defining swimmer orientation whilst affording desirable properties, as it has done here, and will be a subject of future exploration.

In our consideration of a model spermatozoon, the computed yawing function appeared to be approximately sinusoidal, similar to that studied in \cref{sec:trans:sine}. Serendipitously, the yawing function $G(T)=A\sin{T}$ is seen to satisfy the conditions of antiperiodicity and evenness. Hence, the analysis of \cref{sec:drift} suggests that the effects of rapid yawing dynamics are well captured by the leading-order solution on long timescales, and that the approximations derived in this study may be relied upon for significant durations. This simple yawing form also gave rise to unbiased effective dynamics, preserving the direction of swimming and the swimmer's orientation in the context of a Jeffery's orbit, with the sought parity arguments in the general case stemming from our analysis of this minimal example. Further, in this case, the modifications to the swimming speed in the plane and to the Bretherton constant are given analytically in terms of the Bessel function $J_0$, with conversion between shape and swimming parameters in general also being straightforward, given the details of the yawing function. This example additionally highlights the potential for rapid oscillations to significantly alter motion, with certain amplitudes of sinusoidal yawing giving rise to vanishing displacement or even a reversed swimming direction, though we note that large yawing amplitudes are required in order to achieve such drastic changes.

Throughout this work, we have considered simple, phenomenological models of microswimmers in common use, abstracted away from the microscale details of an individual swimmer. Starting with a study of simple planar motion, rendered non-trivial by the presence of rapid yawing, we proceeded to consider scenarios of increasing complexity, incorporating the effects of geometry in flow in \cref{sec:rot} and extending our treatment of translation to three dimensions in \cref{app:3D_trans}. This latter formulation allows for the exploration of three-dimensional yawing, such as the rapid oscillations generated by a model monotrichous bacterium, which would be axisymmetric and, thus, non-yawing were it not for the finite length of its flagellum. Subject to the applicability of a simple model for swimmer motion, this set-up also allows for the consideration of three-dimensional beating of spermatozoa, for instance, a mode of flagellar motion that is frequently exhibited by these canonical swimmers. Returning to planar swimming, as was the focus of the main text and with particular reference to angular dynamics, we have performed high-resolution computational simulations of a model planar-beating spermatozoon in a shear flow in \cref{app:sperm_validity}, evidencing the applicability of an effective Jeffery's orbit model for their rotation in this flow. Such a model has already been employed in other contexts, such as to \leishmex{} by \citet{Walker2018}, though a precise and analytical justification of the suitability of such models remains lacking, and will form the subject of future enquiry.

In summary, we have considered the motion of swimmers and active particles that are subject to rapid oscillations in their orientation. Through asymptotic analysis and supporting numerical explorations, we have quantified how swimming speeds, swimming directions, and even the effective geometries of swimmers and microparticles can be influenced by the presence of high-frequency oscillation, both at leading and sub-leading asymptotic orders. In doing so, we have identified sufficient conditions for oscillations to not introduce directional biases into the emergent dynamics, with the recurring condition of antiperiodic yawing being an observed feature of the motion of a canonical and well-studied swimmer. Finally, in the context of Stokesian fluids, we have seen how the Jeffery's orbits of ellipsoidal and symmetric particles are altered by kinematic oscillations, with the Jeffery's orbit structure of the dynamical system being retained for sinusoidal yawing up to two asymptotic orders, and have highlighted the potential for rapid yawing to enact qualitative long-term changes to the dynamics of bodies in flow.

\section*{Acknowledgements}
The authors are grateful to Prof. Eric Lauga, University of Cambridge, for interesting and motivating discussions on the separated timescales of motion found in many microswimming problems. B.J.W. is supported by the UK Engineering and Physical Sciences Research Council (EPSRC), Grant No. EP/R513295/1. K.I. acknowledges JSPS-KAKENHI for Young Researchers (Grant No. 18K13456) and JST, PRESTO, Japan (Grant No. JPMJPR1921). C.M. is a JSPS International Research Fellow (PE20021). K.I. and C.M. were partially supported by the Research Institute for Mathematical Sciences, an International Joint Usage/ Research Center located at Kyoto University.

\appendix
\section{Translation in three dimensions}\label{app:3D_trans}
In \cref{sec:trans}, we considered the motion of a self-propelled particle in the plane under the influence of prescribed yawing. Here, we extend this leading-order analysis to three dimensions and higher orders, with prescribed average orientation and general yawing that need not be confined to a plane. To do so, we will employ a standard spherical coordinate system to describe the orientation of the swimmer, with $\zeta$ denoting the polar angle and $\phi$ denoting the azimuthal angle, consistent with the notation of the main text. Explicitly, denoting the third spatial coordinate of the swimmer by $z(t)$, the dynamics are governed by the extended dynamical system
\begin{subequations}
\begin{align}
    \diff{x}{t} &= V\sin{\zeta}\cos{\phi}\,,\\
    \diff{y}{t} &= V\sin{\zeta}\sin{\phi}\,,\\
    \diff{z}{t} &= V\cos{\zeta}\,,
\end{align}
\end{subequations}
for fixed speed $V$, noting that $\zeta\equiv\pi/2$ corresponds to the planar system of the main text. As in the general analysis of \cref{sec:trans:general}, we will prescribe the evolution of the swimmer orientation, writing $\zeta = \pi/2 + \Gzeta(T)$ and $\phi = \Gphi(T)$, without loss of generality, for known functions $\Gzeta$ and $\Gphi$ of the fast yawing timescale $T$. As before, these functions completely describe the evolution of the instantaneous swimming direction and may be deduced from simulation or measured from videomicroscopy, for instance. We will assume that both $\Gzeta$ and $\Gphi$ have zero mean over a period in $T$, consistent with choosing a coordinate system in which the average swimmer direction is along the $X$-axis, in line with the forms of $\zeta$ and $\phi$ as written above. Subject to a time-dependent change of coordinates, the analysis of this Appendix may be readily applied if the average heading of the swimmer varies on the slow timescale $t$, though the details of this are notationally cumbersome and are omitted here. Given the reduced setting with a stationary average heading, the dynamical system reads
\begin{subequations}
\begin{align}
    \diff{x}{t} &= V\cos{\Gzeta}\cos{\Gphi}\,,\\
    \diff{y}{t} &= V\cos{\Gzeta}\sin{\Gphi}\,,\\
    \diff{z}{t} &= -V\sin{\Gzeta}\,,
\end{align}
\end{subequations}
from which it is clear that $\Gzeta=\Gphi=0$, i.e. motion with no yawing, corresponds to swimming along the $X$-axis. As in the two-dimensional case, we form the corresponding multiple-scales PDEs and seek expansions for the spatial coordinates in inverse powers of $\omega$. In this case, as in \cref{sec:trans}, the leading-order solutions are independent of the fast timescale $T$, so that we may write $x_0=\bar{x}_0(t)$ etc. These unknown functions correspond to the average motion of the swimmer, with the solvability conditions of the second asymptotic order of the PDEs needed to determine them. These read
\begin{subequations}\label{eq: 3D trans second-order PDEs}
\begin{align}
    \pdiff{x_1}{T} + \diff{\bar{x}_0}{t} &= V\cos{\Gzeta}\cos{\Gphi}\,,\\
    \pdiff{y_1}{T} + \diff{\bar{y}_0}{t} &= V\cos{\Gzeta}\sin{\Gphi}\,,\\
    \pdiff{z_1}{T} + \diff{\bar{z}_0}{t} &= -V\sin{\Gzeta}\,.
\end{align}
\end{subequations}
Seeking periodicity in $T$ and averaging over the fast timescale, we arrive at the leading-order ODEs
\begin{subequations}\label{eq: 3D trans leading-order averages}
\begin{align}
    \diff{\bar{x}_0}{t} &= V\avg{\cos{\Gzeta}\cos{\Gphi}}\,,\\
    \diff{\bar{y}_0}{t} &= V\avg{\cos{\Gzeta}\sin{\Gphi}}\,,\\
    \diff{\bar{z}_0}{t} &= -V\avg{\sin{\Gzeta}}\,.
\end{align}
\end{subequations}
Hence, in order for there to be no directional bias away from the yaw-free heading, which in this coordinate system is along the $X$-axis, we require both 
\begin{equation}
    \avg{\cos{\Gzeta}\sin{\Gphi}} = 0 \quad \text{ and } \quad \avg{\sin{\Gzeta}} = 0\,.
\end{equation}
These conditions are readily found to be satisfied if $\Gzeta$ and $\Gphi$ are both antiperiodic, which naturally extends the conditions of the planar case. As noted in the planar scenario and discussed in more detail in \cref{app:antiperiodic swimmers}, these conditions are automatically satisfied by swimmers with antiperiodic gaits given appropriate definitions of the swimmer orientation, now interpreted to be in three dimensions and including the entire geometry of the body. Should these conditions hold, the effective swimming speed of the self-propelled body is given by $\eff{V} = V\avg{\cos{\Gzeta}\cos{\Gphi}}$, lower in magnitude than the original speed $V$, as is intuitive. For instance, if both $\Gzeta$ and $\Gphi$ were sinusoidal, such as $\Gzeta(T)=\Gphi(T)=A\sin(T)$, then the above conditions are satisfied and the effective velocity is $\eff{V} = V(J_0(2A)+1)/2$.

With the angular dynamics prescribed, the solution at the next order can be easily obtained, as in the case for planar motion. Given the average dynamics of \cref{eq: 3D trans leading-order averages}, we solve \cref{eq: 3D trans second-order PDEs} to give
\begin{subequations}\label{eq: 3D trans second-order solution}
\begin{align}
    x_1 &= \Ix(T) - \avg{\Ix} + \bar{x}_1(t)\,,\\
    y_1 &= \Iy(T) - \avg{\Iy} + \bar{y}_1(t)\,,\\
    z_1 &= \Iz(T) - \avg{\Iz} + \bar{z}_1(t)\,,
\end{align}
\end{subequations}
here defining
\begin{subequations}
\begin{align}
    \Ix &= V\int\limits_0^T \left[\cos{\Gzeta(s)}\cos{\Gphi(s)} - \avg{\cos{\Gzeta}\cos{\Gphi}}\right]\intd{s}\,,\\
    \Iy &= V\int\limits_0^T \left[\cos{\Gzeta(s)}\sin{\Gphi(s)} - \avg{\cos{\Gzeta}\sin{\Gphi}}\right]\intd{s}\,,\\
    \Iz &= -V\int\limits_0^T \left[\sin{\Gzeta(s)} - \avg{\sin{\Gzeta}}\right]\intd{s}\,.
\end{align}
\end{subequations}
Considering the next-order multiple-scales PDEs to determine the unknown averages $\bar{x}_1$, $\bar{y}_1$, and $\bar{z}_1$ and again averaging over $T$, we simply obtain
\begin{equation}\label{eq: 3D trans second-order averages}
    \diff{\bar{x}_1}{t} = 0\,,\quad
    \diff{\bar{y}_1}{t} = 0\,,\quad
    \diff{\bar{z}_1}{t} = 0\,,
\end{equation}
noting that the forcing of the original system of differential equations contributes only at $\bigO{1}$ as the swimmer orientation is prescribed, in contrast to the analysis of \cref{sec:drift}. Hence, the second-order correction to the average swimmer dynamics in three dimensions is simply a constant, so that drift away from this solution appears only in the higher-order terms. These constants are determined by enforcing the initial conditions of $x_1(0,0)=y_1(0,0)=z_1(0,0)=0$ in \cref{eq: 3D trans second-order solution}, giving
\begin{equation}
    \bar{x}_1 = \left.\avg{\Ix}\right|_{t = 0}\,,\quad
    \bar{y}_1 = \left.\avg{\Iy}\right|_{t = 0}\,,\quad
    \bar{z}_1 = \left.\avg{\Iz}\right|_{t = 0}\,.
\end{equation}

\section{An example microswimmer}\label{app:sperm_validity}
\begin{figure}
    \centering
    \begin{overpic}[permil,width = 0.8\textwidth]{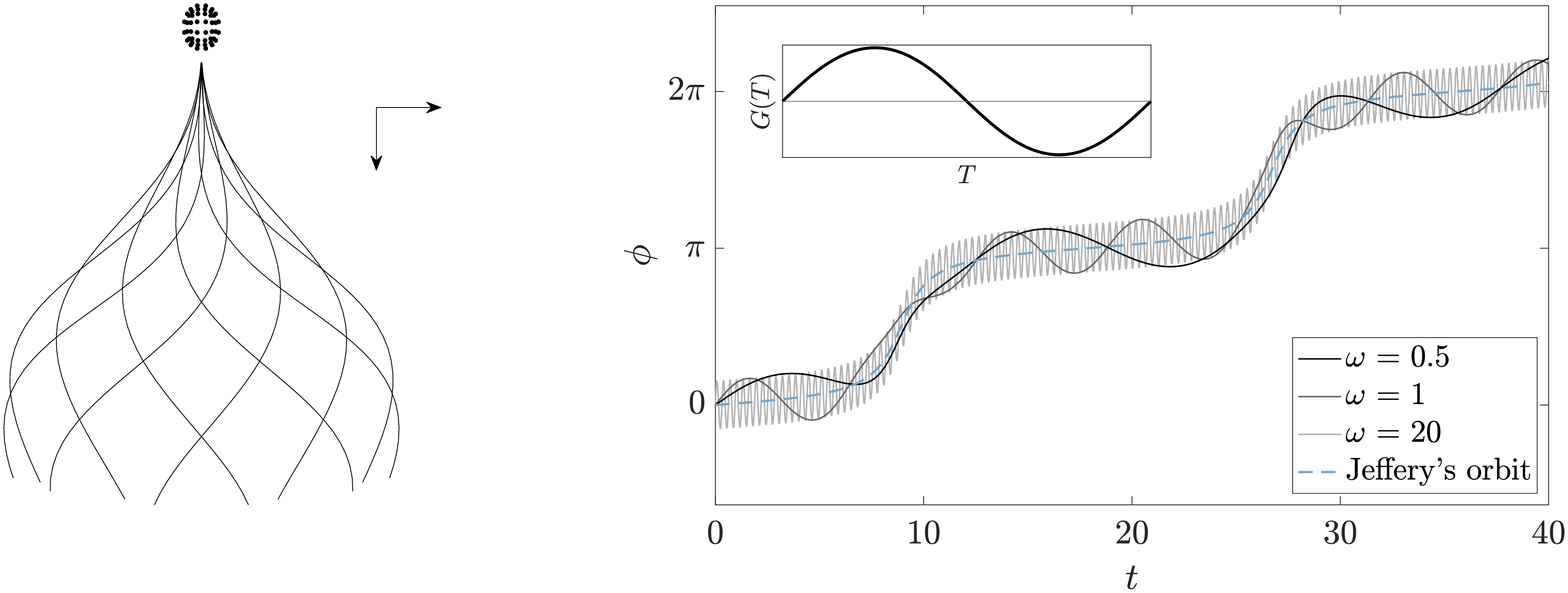}
    \put(-50,380){(a)}
    \put(420,380){(b)}
    \put(214,290){$\hat{x}$}
    \put(247,320){$\hat{y}$}
    \end{overpic}
    \caption{Orbits of a model spermatozoon in shear flow. (a) A computational representation of the model spermatozoon and its beat, shown at multiple points over the period in a swimmer-fixed reference frame. (b) Plotted for $\omega\in\{0.5,1,20\}$, we showcase the evolution of the swimmer orientation $\phi$ as a function of time in a shear flow with $\gamma=1$. We observe a remarkable agreement between the solutions, which lie approximately within an envelope of the temporal moving average, except the lowest frequency case, which is expected from our analysis. Fitting a Jeffery's orbit to the highest frequency case, finding $\eff{B}\approx 0.93573$, we plot this solution as a blue dashed curve, demonstrating the remarkable validity of a Jeffery's orbit model. Inset, we plot the derived yawing function $G(T)$, from which we compute $\trans{z} = 0.9430 - 2\times10^{-4}i$ and $\rot{z} = 0.78151 - 4\times10^{-4}i$, so that both $\trans{\theta}$ and $\rot{\theta}$ are approximately zero and the slow-time motion is unbiased. Further, noting the sinusoidal form of $G$, this yawing function satisfies the conditions of \cref{sec:drift}, so that the long-time accumulation of error occurs at $\bigO{1/\omega^2}$ or higher.}
    \label{fig:sperm_efficacy}
\end{figure}

Though established in \leishmex{} \cite{Walker2018}, the efficacy of Jeffery's orbit approximations for microswimmers in shear flow is far from being universally established. Here, we numerically evidence the validity of utilising a Jeffery's orbit model for the study of human spermatozoa, adopting the morphology of \citet{Gallagher2018a} with the model beat pattern derived from the work of \citet{Dresdner1981}, as shown in \cref{fig:sperm_efficacy}a. Fixing a dimensionless shear rate of $\gamma=1$, we modify the regularised-Stokeslet implementation of \citet{Gallagher2018a} to accommodate the background shear flow, simulating the motion of a beating spermatozoon as it yaws due to the high-frequency flagellar beat. For full details of the methodology and swimmer morphology, we refer the interested reader to the publication of \citet{Gallagher2018a} and the accompanying open-source implementation. In \cref{fig:sperm_efficacy}a, we illustrate the spermatozoan beating gait by plotting snapshots of its flagellum relative to the swimmer body, the latter illustrated by the nodes of a computational mesh and with axis scales shown equal. Here, we have made use of a standard sinusoidal travelling wave for the beating gait, given in terms of a flagellar arclength parameter $s\geq0$ and swimmer-fixed reference coordinates $(\hat{x},\hat{y})$ as
\begin{equation}\label{eq:sperm_gait}
    \hat{x}(s,T) = s\,, \quad \hat{y}(s,T) = (\alpha + \beta s)\sin{\left(ks - T\right)}
\end{equation}
for wavelength $k$ and constants $\alpha=0.1087$ and $\beta=0.0543$ \cite{Dresdner1981}.
From simulation of the swimmer with this gait, temporarily in the absence of background flow and with reference to the notation of the main text, we compute the quantities $\trans{z} = 0.9430 -2\times 10^{-4}i$ and $\rot{z} = 0.78151 - 4\times10^{-4}i$, so that this gait approximately satisfies the conditions of \cref{sec:trans:general} and \cref{sec:drift}. Hence, upon assuming the applicability of the simple models explored in this work, this typical beating pattern would lead to negligible directional bias and is free from accumulating error until at least $\bigO{1/\omega^2}$.

In \cref{fig:sperm_efficacy}b, we plot the evolution of the sperm orientation, defined as the angle $\phi$ between a body-fixed reference frame and a fixed laboratory frame, with the origin of the body frame being at the centre of the sperm head, for multiple values of the angular beat frequency $\omega$. The angle $\phi$ is defined such that $\avg{G}=0$. In all but the lowest-frequency case, and concordant with the presented multiple-scales analysis, the angular evolution of the swimmer can be seen to be well approximated by a fitted Jeffery's orbit, shown as a blue dashed curve with $\eff{B}\approx0.93573$, with the trajectories lying within an envelope of the fitted solution. Further, as the beat frequency of spermatozoa often lies towards the upper range of values considered in this plot, such a Jeffery's orbit approximation is expected to be well suited for modelling the orientation of a beating spermatozoon in a shear flow, though further theoretical justification is required in order to establish this result in more generality.

\section{Antiperiodic swimmers and antiperiodic yaw}\label{app:antiperiodic swimmers}
Here, we detail how swimmers with particular symmetry give rise to antiperiodic yawing in the absence of external influences, focusing on the case where swimmer orientation can be described by a single angle expressed only as a function of the fast time $T$. Consider a swimmer whose geometrical configuration is antiperiodic, by which we mean that the configuration at time $T+\pi$ is a precise reflection (in a fixed plane) of its configuration at time $T$ when viewed in a swimmer-fixed reference frame, such as that shown in \cref{fig:sperm_efficacy}a. Consider also an angle $\phi$ that measures the orientation of the swimmer-fixed frame with respect to a fixed laboratory frame, noting that there are many possible definitions of $\phi$, of which that from \cref{fig:sperm_and_setup}b is an example, and $\phi \mapsto \phi + D$ is another for any constant $D$. The rate of change of $\phi$ with time, written as $\mathrm{d}\phi/\mathrm{d}T$, is easily seen to be antiperiodic, noting the boundary-value nature of the Stokes swimming problem and the assumed antiperiodicity of the boundary conditions and swimmer geometry. This gives rise to a family of derived yawing functions
\begin{equation}
    G(T) = \underbrace{\int\limits_0^T \left.\diff{\phi}{T}\right|_{T=s} \intd{s}}_{G_1(T)} + C\,,
\end{equation}
where this freedom in choosing the constant of integration $C$ is equivalent to shifting $\phi$ by a constant, capturing the many possible definitions of the swimmer orientation. Exploiting the antiperiodicity of the derivative, we can write the yawing functions as
\begin{equation}\label{eq: seeking antiperiodic G}
    G(T) = \left\{\begin{array}{lr}
        G_1(T)+C\,, &  T\in[0,\pi)\,,\\
        G_1(\pi)+C - G_1(T-\pi)\,, & T\in[\pi,2\pi)\,,
    \end{array}\right.
\end{equation}
recalling the $2\pi$-periodicity of $G$ and noting that this satisfies $\mathrm{d}G/\mathrm{d}T = \mathrm{d}\phi/\mathrm{d}T$ for all $T$. From this, it is easy to see that we may ensure that $G$ is antiperiodic by imposing the simple condition $G_1(\pi)+2C=0$, so that $G(T+\pi)+G(T)=0$. Hence, for a swimmer with antiperiodic geometry over the course of its swimming gait, we may readily define its orientation, via $C$, such that its yawing function is antiperiodic. Indeed, this definition is equivalent to choosing $\phi$ such that $\avg{G}=0$, as can be seen directly by computing the average of $G(T)$ as written in \cref{eq: seeking antiperiodic G}, finding that this average vanishes if and only if $G_1(\pi)+2C=0$. 

Of note, the above analysis extends to the antiderivatives of any antiperiodic function, with the antiderivative of an antiperiodic function being antiperiodic if and only if the average of the antiderivative is zero over a full period, such as that shown in \cref{fig:G_examples}b.

\section{Numerically exploring the effects of frequency on error}\label{app:error_scaling}
\begin{figure}[ht]
    \centering
    \includegraphics[width=0.4\textwidth]{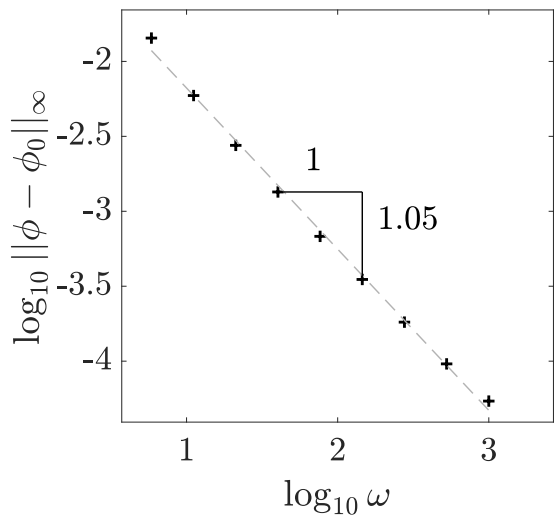}
    \caption{Error in the multiple-scales solution for planar Jeffery's orbits with sinusoidal yawing. For a range of values of $\omega$, we numerically solve the perturbed system \cref{eq:full_system_phi}, comparing the solution with that of the multiple-scales analysis of \cref{eq:leading_order_ODE_phi0}. We plot, as black crosses, the base ten logarithm of the infinity norm error in the orientation as a function of $\log_{10}{\omega}$, clearly observing the predicted $1/\omega$ behaviour as $\omega\rightarrow\infty$. Here, the dynamics are simulated over ten periods of the unperturbed Jeffery's orbit, taking $B=0.5$, $\gamma=1$, and $A=0.1$. A linear fit is shown as a grey, dashed line.}
    \label{fig:error_scaling}
\end{figure}

To support the multiple-scales analysis for the planar Jeffery's orbit in the presence of sinusoidal yawing, we numerically solve the perturbed system of \cref{eq:full_system_phi} and compare the solution to the leading-order multiple-scales approximation of \cref{eq:leading_order_ODE_phi0}. Computation was performed using the in-built ordinary differential equation solver \texttt{ode45} in MATLAB \cite{Shampine1997} with error tolerances of $10^{-9}$. In \cref{fig:error_scaling}, we plot the base ten logarithm of the infinity norm error between the full solution $\phi$ and the leading-order multiple-scales solution $\phi_0$, denoted by $\log_{10}{\norm{\phi - \phi_0}_{\infty}}$, against $\log_{10}{\omega}$, comparing the solutions over ten periods of the unperturbed Jeffery's orbit. Immediately evident is the approximate $1/\omega$ scaling of the error with the frequency of oscillation, in line with the analysis of \cref{sec:rot} and confirming the predicted accuracy of the multiple-scales approximation over this timescale. In particular, this order of error reflects the difference between the initial conditions of the leading-order and higher-order approximation, which scale with $1/\omega$.

\section{Details of higher-order drift}
\label{app:drift_details}
In the analysis of \cref{sec:drift}, we remarked that agreement between the average leading-order dynamics of \cref{eq: leading_order drift ODEs} and the higher-order averaged dynamics of \cref{eq:joint drift} required that $\trans{\znew} - \trans{z} = \smallO{1/\omega}$ and $\rot{\znew} - \rot{z} = \smallO{1/\omega}$. Given the leading-order agreement, it remains to show that the sub-leading terms of $\trans{\znew}$ and $\rot{\znew}$ vanish, subject to constraints on the yawing function $G$. Expanding out the exponential integrands of the averages that define these complex numbers in \cref{eq: modified za zp} and retaining only the relevant terms, we have
\begin{subequations}
\label{eq: znew minus z}
\begin{align}
    \trans{\znew} - \trans{z} &= -\frac{1}{\omega}\avg{\left[\Iphi - \avg{\Iphi}\right]\sin{G} - i\left[\Iphi - \avg{\Iphi}\right]\cos{G}} + \smallO{1/\omega}\,,\\
    \rot{\znew} - \rot{z} &= -\frac{2}{\omega}\avg{\left[\Iphi - \avg{\Iphi}\right]\sin{2G} - i\left[\Iphi - \avg{\Iphi}\right]\cos{2G}} + \smallO{1/\omega}\,.
\end{align}
\end{subequations}
Hence, we must compute the averages of terms that involve the quantity $\Iphi - \avg{\Iphi}$ multiplied by trigonometric functions of $G$ and $2G$. To do so, it is helpful to recall the definition of $\Iphi$ from \cref{eq: I definitions}, which can be expanded as 
\begin{subequations}
\begin{align}
    \Iphi(t,T) = \dfrac{\gamma B}{2} \left[\cos 2 \bar{\phi}_0(t) \Ic(T) - \sin 2 \bar{\phi}_0(t) \Is(T) \right]\,,
\end{align}
where
\begin{align}
\label{eq: Ics definitions}
\Ic(T) \coloneqq \int\limits_0^{T} \! \left[\avg{\cos 2 G} - \cos 2 G(s)\right] \intd{s}, \quad
\Is(T) \coloneqq \int\limits_0^{T} \! \left[\avg{\sin 2 G} - \sin 2 G(s)\right] \intd{s},
\end{align}
\end{subequations}
utilising the general expression for the leading-order angular solution $\phi_0$ of \cref{eq: phi_0 general solution}. Thus, $\Iphi$ is a linear combination of $\Ic$ and $\Is$, with coefficients that are independent of $T$, so that $\Iphi-\avg{\Iphi}$ is a linear combination of $\Ic - \avg{\Ic}$ and $\Is - \avg{\Is}$. Hence, the terms of \cref{eq: znew minus z} will vanish if we can demonstrate that the following eight terms each have zero average over a period in $T$:
\begin{subequations}\label{eq: list of terms to have zero mean}
\begin{align}
    \left[\Ic - \avg{\Ic}\right]\sin{G}\,, &&\quad \left[\Ic - \avg{\Ic}\right]\cos{G}\,, &&\quad \left[\Is - \avg{\Is}\right]\sin{G}\,, &&\quad \left[\Is - \avg{\Is}\right]\cos{G}\,,\\
    \left[\Ic - \avg{\Ic}\right]\sin{2G}\,, &&\quad \left[\Ic - \avg{\Ic}\right]\cos{2G}\,, &&\quad \left[\Is - \avg{\Is}\right]\sin{2G}\,, &&\quad \left[\Is - \avg{\Is}\right]\cos{2G}\,.
\end{align}
\end{subequations}
Here, the first row of terms pertains to the translational motion, whilst the second row concerns the angular evolution. We will show that $\left[\Ic - \avg{\Ic}\right]\cos{2G}$ and $\left[\Is - \avg{\Is}\right]\sin{2G}$ have identically zero averages in all cases, whilst the remaining terms have vanishing average due to parity arguments, following the imposition of two appropriate symmetry conditions on $G$. 

We begin with the two terms that do not require additional conditions on $G$. Instead of $\left[\Ic - \avg{\Ic}\right]\cos{2G}$, consider the modified expression $\left[\Ic - \avg{\Ic}\right]\left[\avg{\cos{2G}} - \cos{2G}\right]$, which has the same average up to sign as the original by virtue of $\avg{\Ic - \avg{\Ic}}=0$. Recalling the definition of $\Ic$ and defining $F(T) = \Ic - \avg{\Ic}$, so that $F'(T) = \avg{\cos{2G}} - \cos{2G(T)}$, we see that we can write
\begin{equation}
    \avg{\left[\Ic - \avg{\Ic}\right]\left[\avg{\cos{2G}} - \cos{2G}\right]} = \frac{1}{2\pi}\int\limits_0^{2\pi}FF'\intd{T} = \frac{1}{4\pi}\left[F^2\right]_0^{2\pi} = 0
\end{equation}
as $F$ is periodic, a property inherited from $G$. Hence, the average of $\left[\Ic - \avg{\Ic}\right]\cos{2G}$ is zero. Analogous reasoning applies to $\left[\Is - \avg{\Is}\right]\sin{2G}$, so that it also has zero mean, with no conditions on $G$ necessary.

Next, we impose an antiperiodicity condition on $G$; precisely the condition that was demonstrated to remove the leading-order orientational biases in the translational and angular dynamics in \cref{sec:trans:general} and \cref{sec:rot}, respectively. This antiperiodicity immediately means that $\sin{G}$ and $\sin{2G}$ are also antiperiodic, whilst $\cos{G}$ and $\cos{2G}$ are periodic with period $\pi$. From this, we will now seek to establish the antiperiodicity of $\Is - \avg{\Is}$ and the $\pi$-periodicity of $\Ic - \avg{\Ic}$, which will lead to the vanishing of all but two of the remaining terms in \cref{eq: list of terms to have zero mean} by parity. 

In fact, the $\pi$-periodicity of $\Ic$, and therefore $\Ic - \avg{\Ic}$, follows immediately from its integrand $\avg{\cos{2G}} - \cos{2G}$ being $\pi$-periodic and having zero mean over a period in $T$. The $\pi$-antiperiodicity of $\Is - \avg{\Is}$ follows from the concluding remarks of \cref{app:antiperiodic swimmers}, with its integrand $\avg{\sin{2G}} - \sin{2G}$ being $\pi$-antiperiodic, and having zero mean over a $2\pi$-period in $T$.

The above parity arguments, which assume only that $G$ is antiperiodic, are sufficient to render the average of all of the remaining terms of \cref{eq: list of terms to have zero mean} zero, with the exception of the translational terms $\left[\Ic - \avg{\Ic}\right]\cos{G}$ and $\left[\Is - \avg{\Is}\right]\sin{G}$. Hence, we can conclude that the antiperiodicity of $G$ is sufficient to ensure that the average leading-order \emph{angular} dynamics are equal to the higher-order averaged angular dynamics, with any deviation occurring at $\bigO{1/\omega^2}$ or higher.

An additional condition is required in order to draw the same conclusion of the translational dynamics. Here, once again seeking to make use of a parity argument, we impose the additional constraint that $G$ is even about some $T_e^{\star}$, which we note is satisfied by sinusoidal yawing. This implies that both $\cos{G}$ and $\sin{G}$ are even about the same $T_e^{\star}$, whilst also ensuring that $\Ic - \avg{\Ic}$ and $\Is - \avg{\Is}$ are odd about $T_e^{\star}$. These latter conclusions require some exposition. Consider $\Ic - \avg{\Ic}$, and write
\begin{equation}\label{eq: Ic even}
    \Ic(T_e^{\star} + T) - \avg{\Ic} = \left(\int_0^{T_e^{\star}} + \int_{T_e^{\star}}^{T_e^{\star} + T}\right)\left[\avg{\cos{2G}} - \cos{2G(s)}\right]\intd{s} - \avg{\Ic}\,.
\end{equation}
The second integral in the above is odd in $T$, with both $\avg{\cos{2G}}$ and $\cos{2G}$ being even about $T_e^{\star}$. Hence, averaging over $T$, we have
\begin{equation}
    0 = \int\limits_0^{T_e^{\star}}\left[\avg{\cos{2G}} - \cos{2G(s)}\right]\intd{s} - \avg{\Ic}\,.
\end{equation}
Inserting this relation into \cref{eq: Ic even} gives
\begin{equation}
    \Ic(T_e^{\star} + T) - \avg{\Ic} = \int_{T_e^{\star}}^{T_e^{\star} + T}\left[\avg{\cos{2G}} - \cos{2G(s)}\right]\intd{s}\,,
\end{equation}
which we have already noted is odd in $T$. Hence, $\Ic-\avg{\Ic}$ is odd about $T_e^{\star}$. An entirely analogous analysis yields that $\Is-\avg{\Is}$ is odd about $T_e^{\star}$.

Hence, by parity, the remaining terms of \cref{eq: list of terms to have zero mean} have vanishing average, and we can conclude that, if $G$ is both antiperiodic and even, the average leading-order dynamics of \cref{eq: leading_order drift ODEs} are equal to the higher-order averaged dynamics of \cref{eq:joint drift}, with corrections arising at $\bigO{1/\omega^2}$ or higher.

\end{document}